\pdfoutput=1
\documentclass[iop,apj]{emulateapj}

\def\lessim{\mathrel{\hbox{\rlap{\hbox{\lower4pt\hbox{$\sim$}}}\hbox{$<$}}}}
\def\grtsim{\mathrel{\hbox{\rlap{\hbox{\lower4pt\hbox{$\sim$}}}\hbox{$>$}}}}

\begin{document}

\title{The Galactic Nova Rate Revisited}
\author{A. W. Shafter\altaffilmark{1}}
\altaffiltext{1}{Department of Astronomy and Mount Laguna Observatory, San Diego State University, San Diego, CA 92182}

\begin{abstract}
Despite its fundamental importance,
a reliable estimate of the Galactic nova rate has
remained elusive. Here, the
overall Galactic nova rate is estimated by extrapolating
the observed rate for novae reaching $m\leq2$
to include the entire Galaxy
using a two component disk plus bulge
model for the distribution of stars in the Milky Way.
The present analysis improves on previous work by considering
important corrections for incompleteness in the observed rate of
bright novae and by employing a Monte Carlo analysis to better estimate
the uncertainty in the derived nova rates.
Several models are considered to account for
differences in the assumed properties of bulge and disk nova populations
and in the absolute magnitude distribution.
The simplest models, which assume uniform properties between
bulge and disk novae, predict Galactic nova rates of $\sim$50
to in excess of 100 per year,
depending on the assumed incompleteness
at bright magnitudes.
Models where the
disk novae are assumed to be more luminous than bulge novae
are explored, and predict nova
rates up to 30\% lower,
in the range of $\sim$35 to $\sim$75 per year.
An average of the most plausible models yields a rate of
$50_{-23}^{+31}$~yr$^{-1}$, which is arguably the
best estimate currently available for the nova rate in the Galaxy.
Virtually all models produce rates that
represent significant increases over recent estimates,
and bring the Galactic nova rate into better agreement with that expected
based on comparison with the latest results from extragalactic surveys.
\end{abstract}


\section{Introduction}

The Galactic nova rate is important in the study of a variety of
astrophysical problems.
For example,
classical nova explosions are thought to play a significant role in the
chemical evolution of the Galaxy \citep[e.g.,][and references therein]{jos06}.
In addition to producing a fraction of the
$^7$Li and the short-lived isotopes $^{22}$Na and $^{26}$Al,
novae are believed to be important in the production of the CNO isotopes,
particularly $^{15}$N, where novae may account for
virtually all of the Galactic abundance of this isotope.
Thus, complete models for Galactic chemical evolution necessarily
require the nova rate as an input parameter.

Novae may also play an important role as Type Ia supernova (SN~Ia) progenitors
\citep[][and references therein]{sha15,sor15,sta16}.
Indeed, perhaps the most promising SN~Ia progenitor extant
is the recurrent nova, M31N 2008-12a \citep{hen15a,tan14,dar15}.
M31N 2008-12a has an extremely short recurrence time of just under a year,
possibly as short as 6 months \citep{hen15b}, which constrains the
accretion rate to be $2-3\times10^{-7}$~M$_{\odot}$~yr$^{-1}$ and the mass
of the white dwarf to be near the Chandrasekhar limit \citep{kat14,wol13}.
These models also suggest that the white dwarf
is gaining mass, and that it
will reach the Chandrasekhar limit in less than $10^6$ years.
The ultimate fate of M31N 2008-12a, as with all similar recurrent nova systems,
depends on whether the composition of the white dwarf is CO or ONe. In the
former case the system is expected to explode as a SN~Ia, while in the latter
case electron captures onto $^{20}$Ne and $^{24}$Mg
will result in an accretion-induced collapse
and the subsequent formation of a neutron star \citep{miy80}.

Despite its importance, the Galactic nova rate is not well established.
Estimates have varied widely, from as few as 20
to as many as 260~yr$^{-1}$ \citep{del94, sha72}.
Recently, \citet{mro15} have measured a rate of $13.8\pm2.6$~yr$^{-1}$ for the
Galactic bulge alone based on OGLE observations.
Global nova rates
have been estimated both directly,
by extrapolating the observed rate in the vicinity of the
sun to the entire Galaxy \citep[e.g., see][]{sha97, sha02}, and indirectly
through comparison with other galaxies \citep{sha00,dar06,sha14}.
\citet{sha02} used the \citet{bah80}
model for the stellar density in the Milky Way to extrapolate the
rate of novae with $m\lessim2$, which was assumed to be complete,
to faint magnitudes
finding a Galactic rate of $36\pm13$~yr$^{-1}$. Recently, \citet{she14}
performed a thorough analysis of the observational selection biases against the
discovery of even the brightest novae. After taking these
biases into account, \citet{she14} makes the surprising suggestion
that only $43\pm6$\% of Galactic novae with $m\leq2$ are likely recovered.

In this paper, we reconsider the
analysis presented in \citet{sha02} and \citet{sha97}
by conducting Monte Carlo simulations to estimate the global Galactic
nova rate and its uncertainty. The revised analysis
includes the increased sample of Galactic novae available since 2000,
and makes more plausible assumptions regarding
the completeness of the nova sample at bright magnitudes.
We conclude by comparing the latest Galactic nova estimates with those
recently measured in extragalactic systems such as the nearby spiral M31, and
the Virgo elliptical M87.

\begin{figure*}

\includegraphics[scale=0.66,angle=-90]{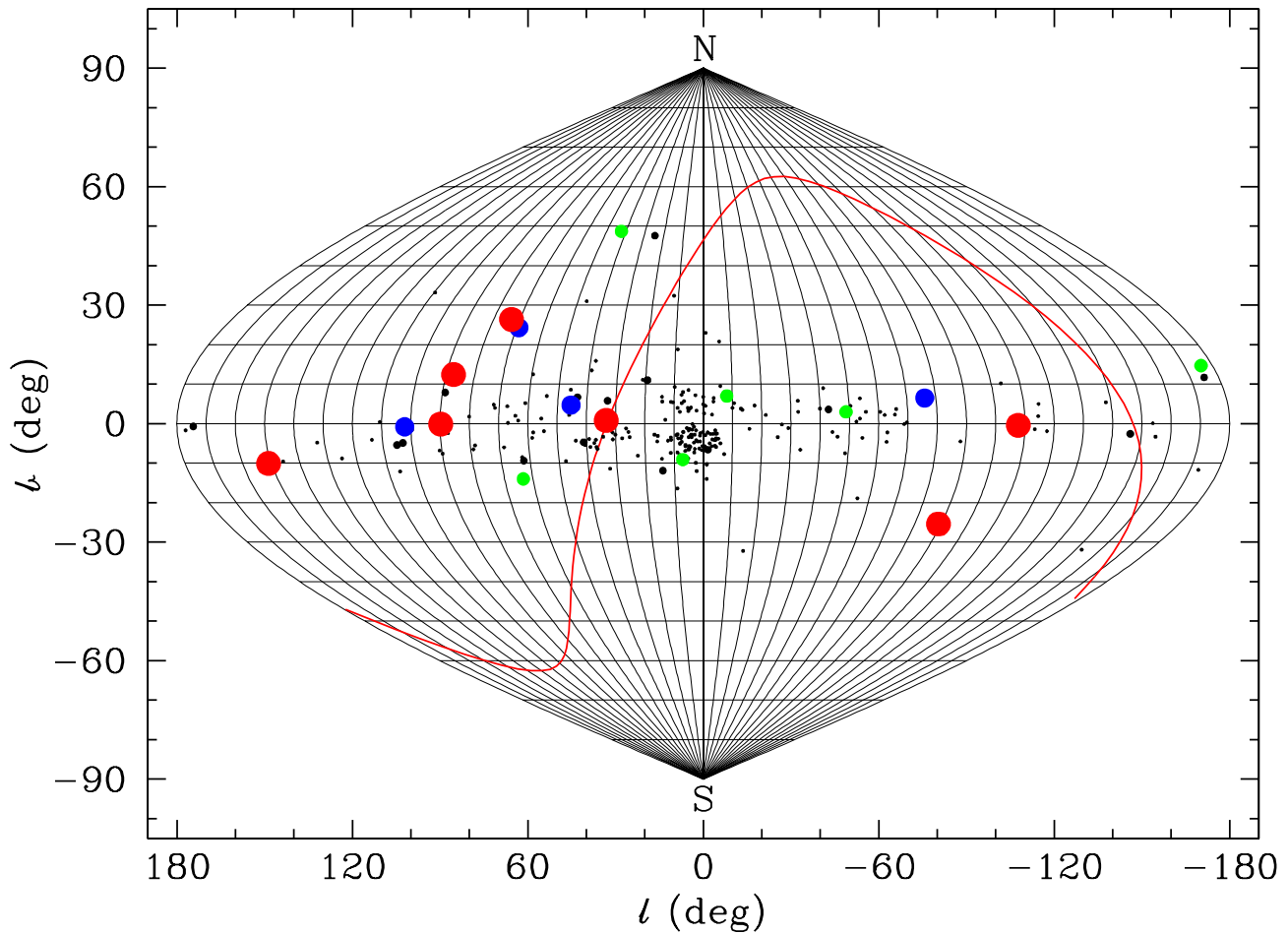} 

\caption{The distribution of Galactic novae brighter than $m=10$ discovered
since 1900 are displayed in Galactic coordinates.
Key -- red filled circles: $m\leq2$, smaller blue filled circles: $m\leq3$,
smaller green filled circles: $m\leq4$, smaller black filled circles: $m\leq5$,
smallest black dots: $m\leq10$.
The solid red line
represents the Celestial equator. Note the large concentration of faint
novae toward the Galactic center and the apparent
bias for brighter novae in the Northern hemisphere.
\label{fig1}}

\end{figure*}

\section{The Observed Nova Sample}

From 1900 to the end of 2015
there have been a total of $\sim$250 Galactic novae discovered brighter than
$m=10$, and we have compiled a list of these novae in Table~\ref{tab1}.\footnote{
Because Galactic nova observations have been variously reported as
photographic ($pg$), $B$, and $V$ magnitudes, we make no attempt to
correct to a common effective wavelength.
The fact that novae shortly after maximum light have $B-V\simeq0$
\citep{van87} suggests that any corrections would not be significant.}
The spatial positions of the full nova sample from Table~\ref{tab1} are
plotted in Figure~\ref{fig1}, which confirms that the novae are 
concentrated in the Galactic plane, and in the direction of the Galactic center.
The observed asymmetry about Galactic latitude $b=0$ confirms the presence of
significant and patchy absorption in the disk plane. Clearly,
magnitude-limited samples of novae will be incomplete,
particularly at fainter magnitudes.

\begin{deluxetable*}{lrlrrrr}
\tablenum{1}
\label{tab1}
\tablewidth{0pt}
\tablecolumns{7}
\tablecaption{Galactic Novae Since 1900}
\tablehead{\colhead{Nova} & \colhead{Name} & \colhead{Date (Y,M,D)} & \colhead{$l$ ($^{\circ}$)} & \colhead{$b$ ($^{\circ}$)} & \colhead{$m_{dis}$\tablenotemark{a}} &  \colhead{Ref\tablenotemark{b}} }

\startdata

N Sgr 2015\#3   & V5669 Sgr & 2015   9  27 & $   3.0$ & $ -2.8$ &  8.7 & 1 \\
N Sgr 2015\#2   & V5668 Sgr & 2015   3  15 & $   7.1$ & $ -9.1$ &  4.0 & 1 \\
N Sgr 2015\#1   & V5667 Sgr & 2015   2  12 & $   7.4$ & $ -3.2$ &  9.0 & 1 \\
N Sco 2015\#1   & V1535 Sco & 2015   2  11 & $ 350.0$ & $  4.0$ &  8.2 & 1 \\
N Sgr 2014     & V5666 Sgr & 2014   1  26 & $  11.0$ & $ -4.1$ &  8.7 & 1 \\
N Cen 2013     & V1369 Cen & 2013  12   2 & $ 311.1$ & $  3.0$ &  3.3 & 1 \\
N Del 2013     & V339 Del  & 2013   8  14 & $  62.2$ & $ -9.4$ &  4.3 & 1 \\
N Cep 2013     & V809 Cep  & 2013   2   2 & $ 110.6$ & $  0.4$ &  9.8 & 1 \\

\enddata
\tablenotetext{a}{Peak magnitude when available, otherwise discovery magnitude}
\tablenotetext{b}
{(1) {\tt http://asd.gsfc.nasa.gov/Koji.Mukai/novae/novae.html} }
\tablecomments{Table~\ref{tab1} is published in its entirety in a machine readable
format. A portion is shown here for guidance regarding its form and content.}

\end{deluxetable*}

\begin{figure}

\includegraphics[scale=0.345]{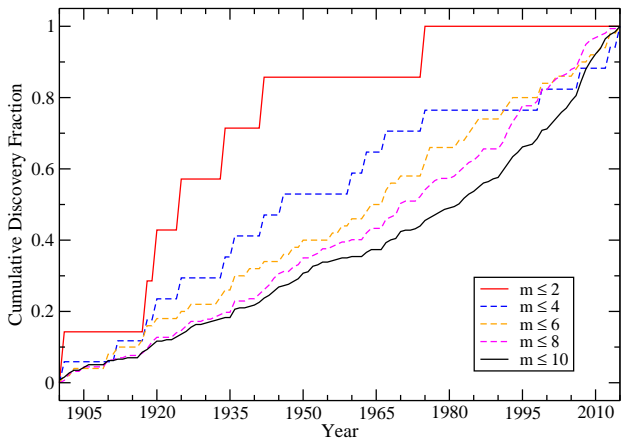} 

\caption{Cumulative distribution of Galactic novae discovery magnitudes.
Despite the expected increase in the discovery rate for fainter novae
in recent years,
note the striking drop in the discovery of bright novae since $\sim$1950.
\label{fig2}}

\end{figure}

\begin{figure}

\includegraphics[scale=0.345]{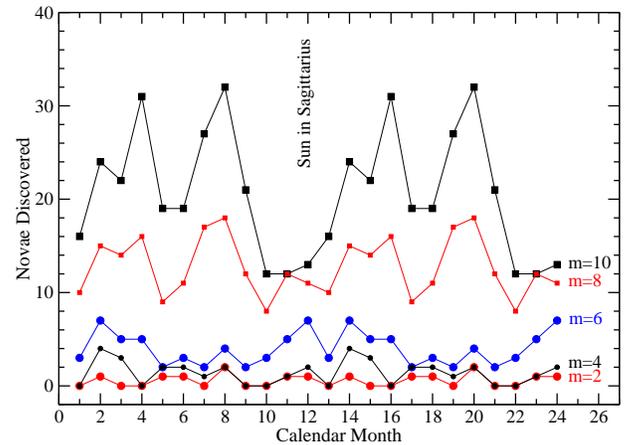} 

\caption{The number of novae discovered since 1900 plotted as a function
of the month of discovery. At the faintest magnitudes a drop in the
discovery rate is apparent when the sun is Sagittarius, significantly
diminishing the discovery of novae in the direction of the Galactic center.
\label{fig3}}

\end{figure}

Surprisingly,
novae at the brightest magnitudes appear to be biased toward the northern
hemisphere. In Table~\ref{tab2} we show the number of bright novae ($m\le5$) discovered
north and south of the celestial equator since 1900 as a function
of apparent magnitude.
At first glance it appears that we
may be missing a significant fraction of novae at brighter magnitudes
in the southern hemisphere, but the observed sample
of novae is small.
We can gain some insight into the statistical significance of this
apparent asymmetry as follows. For a given magnitude,
the probability that $N$ or fewer
novae would be discovered in a given hemisphere out of a total of
$M$ novae is given by:

\begin{equation}
P_{\leq N,M} =  \sum_{N=0}^N {M! \over N! (M - N)!} \  0.5^M
\end{equation}

Since the excess could have occurred in either hemisphere, we must
multiply the probability given in equation~(1) by two.
For $m=2$ we have $N=2$ and $M=7$, which results in $P_{\leq 2,7} = 0.45$.
Thus, the fact that only two of the seven novae brighter than $m=2$ have
erupted in the southern hemisphere is not particularly surprising.
Probabilities for $m=3, 4$ and 5 are also given in Table~\ref{tab2}. It seems clear
that given the small number statistics, we cannot claim any statistically
significant bias in bright nova detections to any one hemisphere.

\begin{deluxetable}{ccccccc}
\tabletypesize{\scriptsize}
\tablenum{2}
\label{tab2}
\tablewidth{0pt}
\tablecolumns{7}
\tablecaption{Equitorial Distribution of Bright Nova Discoveries}
\tablehead{\colhead{$m_\mathrm{lim}$} & \colhead{$N_{\mathrm south}$} & \colhead{$N_{\mathrm north}$} & \colhead {$N_{\mathrm tot}$}& \colhead{$P$}}

\startdata
 2    &    2    &    5    & 7   & 0.45\\
 3    &    3    &    8    & 11  & 0.23 \\
 4    &    6    &   11    & 17  & 0.33 \\
 5    &   12    &   21    & 33  & 0.16 \\

\enddata

\end{deluxetable}

\begin{deluxetable*}{llcccccr}
\tablenum{3}
\label{tab3}
\tablewidth{0pt}
\tablecolumns{8}
\tablecaption{Galactic Novae with $m\leq2$}
\tablehead{\colhead{Name} & \colhead{Date (Y,M,D)} & \colhead{$l$ ($^{\circ}$)} & \colhead{$b$ ($^{\circ}$)} & \colhead{$m_{dis}$\tablenotemark{a}} &  \colhead{d (kpc)} & \colhead{$E_{B-V}$ (mag)} & \colhead{Ref\tablenotemark{b}} }

\startdata

V1500 Cyg & 1975   8  29 & $  89.8$ & $ -0.1$ &  1.9 & $1.5\pm0.2$&$0.43\pm0.08$& 1,2 \\
CP Pup    & 1942  11   9 & $252.3$ & $ -0.4$ &  0.5 & $1.5$ &$0.25\pm0.06$& 1,2 \\
DQ Her    & 1934  12  12 & $  73.2$ & $ 26.4$ &  1.3 & $0.40\pm0.06$&$0.08$&1,2 \\
RR Pic    & 1925   5  25 & $ 271.0$ & $-25.4$ &  1.0 & 0.4 & 0.02& 1,2 \\
V476 Cyg  & 1920   8  20 & $  87.4$ & $ 12.4$ &  2.0 & $1.62\pm0.12$&$0.27\pm0.12$&1,2 \\
V603 Aql  & 1918   6   8 & $  33.2$ & $  0.8$ & $-1.1$ & 0.33 & 0.08 & 1,2 \\
GK Per    & 1901   2  21 & $ 151.0$ & $-10.1$ &  0.2 & $0.46\pm0.0.03$& 0.29 & 1,2 \\

\enddata
\tablenotetext{a}{Peak magnitude when available, otherwise discovery magnitude}
\tablenotetext{b}
{(1) {\tt http://asd.gsfc.nasa.gov/Koji.Mukai/novae/novae.html};
 (2) \citet{dar12}.}

\end{deluxetable*}

Nevertheless, evidence for possible incompleteness at
bright magnitudes can be appreciated by considering how the rate of
discovery of novae reaching apparent magnitude $m$,
$N(m)$, has varied over time.
Figure~\ref{fig2} shows the cumulative distribution
of nova discovery magnitudes during the period between 1900 and 2015.
Of the seven novae that reached $m=2$ or brighter,
six were discovered in the first half (58 yrs) of this period
(see Table~\ref{tab3}).
Only one nova, Nova Cyg 1975 (V1500 Cyg),
which reached $m=1.9$, has been discovered in the second 58
year interval (actually, in the last 73 years!).
We can compute the significance of this result by computing
the probability that $N$ or fewer novae with $m\leq2$ would be
found within any consecutive 58 year span in the 116 years since 1900.
That probability
is given by equation~(1), where
$M-1$ novae must erupt within the same 58 year window.
In this case, where $N=1$, we find
$P = 6\times0.5^6~(\rm{one~nova})~+ 0.5^6~(\rm{no~novae}) = 0.11$.
Assuming that the true nova rate has been constant over time,
a KS test reveals a similar result, namely
only a 7\% chance that novae
with $m\leq2$ would be distributed as shown in Figure~\ref{fig2}.
Despite the fact that these probabilities do not rule out
100\% completeness for $m\leq2$ at the $2\sigma$ level,
the probabilities are small, and suggest that
at least one nova was likely missed in recent years.
With only seven of a possible eight $m\leq2$ novae being detected
since 1900, the completeness becomes $\sim88$\%.

Although it appears counterintuitive, \citet{she14} points out
that incompleteness at brighter magnitudes may be due in part
to the evolution of how amateur astronomers survey the sky, which in recent
years has turned to the use of telescopes equipped with CCD detectors
rather than memorizing the sky and conducting wide-area visual observations.
In addition, seasonal effects (i.e., sun in Sagittarius) will
also have diminished the observed rate of fainter novae
concentrated towards the Galactic center (see Figure~\ref{fig3}).
As mentioned earlier,
after considering a variety of selection effects \citet{she14} arrived
at a completeness of just 43\% for novae brighter than $m=2$.
Because only a summary of this work has been published, it is not possible
to critically evaluate the assumptions made in arriving at this value.
It does, however, seem quite surprising that more than half of the novae
reaching second magnitude since 1900 could have been missed.
Whether the completeness is close to 90\% as estimated above, or whether
Schaefer is correct that we have missed more than half of the second magnitude
and brighter novae, one thing seems clear,
the assumption of 100\% completeness for $m\leq2$
made earlier by \citet{sha02} is likely to be overly optimistic.

In the analysis to follow, we constrain the Galactic nova rate
by adopting plausible limits on the completeness of bright novae.
Given that
it seems difficult to understand how the completeness could be lower
than the value determined by \citet{she14},
we have adopted $c=0.43$ as a lower limit
on the completeness of novae with $m\leq2$.
The possibility that all novae with $m\leq2$ have been detected
since 1900 provides a hard upper limit of 100\% on the completeness.
We argue that the best estimate of the completeness lies between these limits,
and follows from two considerations.
As described earlier, the sharp drop in the number of $m\leq2$ novae
observed over the past $\sim$60~yr suggests that at least one
out of eight bright novae has likely been missed in recent years.
If so, a value of $c=0.88$ would seem
to offer a reasonable estimate for the completeness of novae with $m\leq2$.
This estimate is supported by considering that
even the brightest novae will likely be missed
if they erupt within $18^{\circ}$ (1.2 hr) of the sun. Based on this
correction
alone, the completeness drops to $\sim90$\%.
Thus, in computing the models described in the following section,
we simply take $c=0.9$ as our best
estimate of the completeness for novae with $m\leq2$.
For comparison, we also consider models for 
$c=0.43$, which we take as a
lower limit to the
completeness of novae reaching second magnitude or brighter.

\begin{figure}

\includegraphics[scale=0.345]{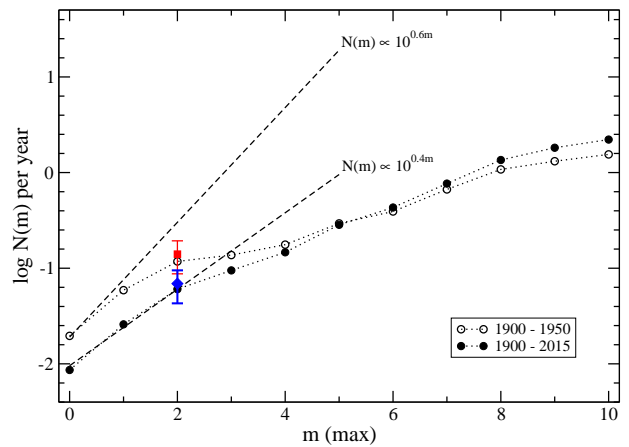} 

\caption{Cumulative distribution of Novae as a function of peak magnitude.
Note the higher average nova rate for novae with $m\lessim2$ during
the period between $1900-1950$ compared with the period from 1900 to
the present.
The red square and blue diamond
represent the value of log $N(2)$ for the full $1900-2015$ interval
corrected for
the $c=0.43$ completeness of \citet{she14}, and for our $c=0.9$ completeness
estimate, respectively.
The error bars represent the Poisson error for the seven novae
discovered with $m\leq2$.
The dashed lines represents the expected increase in the
nova rate for an infinite uniform distribution of nova progenitors
(log $N \propto 0.6m$),
and for an infinite disk distribution (log $N \propto 0.4m$).
\label{fig4}}

\end{figure}

\section{Model}

The annual discovery rate of novae brighter than $m=10$ since 1900
as a function of magnitude is shown in Figure~\ref{fig4}.
The corrected $m\leq2$ nova discovery rates
for our estimated completeness of $c=0.9$ and the lower completeness
advocated by \cite{she14}
are shown as the blue diamond and the red square, respectively.
For comparison, we have also plotted the values of $N(m)$ computed from a
sub-sample of novae discovered during the period between
1900 and 1950. We find that the annual discovery rate for novae
reaching second magnitude or brighter during this earlier time span
is approximately twice that for the
full $1900 - 2015$ period, and
is consistent with that expected after
applying Schaefer's incompleteness estimate.

In the analysis to follow, we estimate
the Galactic nova rate by
extrapolating the local nova rate
($m\leq2$) to the entire Galaxy
based on a model consisting of separate
bulge and disk components.
The bulge component, $\rho_b$,
is modeled using a standard \cite{dev59} luminosity profile, while
The disk nova density, $\rho_d$,
is assumed to
have a double exponential dependence on distance from the Galactic center and
on the distance from the Galactic plane.

\begin{deluxetable}{lrrrc}
\tablenum{4}
\label{tab4}
\tablewidth{0pt}
\tablecolumns{5}
\tablecaption{Distances of Novae from the Galactic Plane}
\tablehead{\colhead{Nova} & \colhead{$d$ (kpc)} & \colhead{$b$ ($^{\circ}$)} & \colhead{$z$ (kpc)} & \colhead{Ref\tablenotemark{a}} }

\startdata
CI Aql      &   5.00 &$  -0.8$&  0.070 &   1  \\
V356 Aql    &   1.70 &$  -4.9$&  0.145 &   1  \\
V528 Aql    &   2.40 &$  -5.9$&  0.247 &   1  \\
V603 Aql    &   0.25 &$   0.8$&  0.003 &   2  \\
V1229 Aql   &   1.73 &$  -5.4$&  0.163 &   1  \\
T Aur       &   0.96 &$  -1.7$&  0.028 &   1  \\
IV Cep      &   2.05 &$  -1.6$&  0.057 &   1  \\
V394 CrA    &  10.00 &$  -7.7$&  1.340 &   1  \\
T CrB       &   0.90 &$  48.2$&  0.671 &   1  \\
V476 Cyg    &   1.62 &$  12.4$&  0.348 &   1  \\
V1500 Cyg   &   1.50 &$  -0.1$&  0.003 &   1  \\
V1974 Cyg   &   1.77 &$  -9.6$&  0.295 &   1  \\
V2491 Cyg   &  13.30 &$   4.4$&  1.020 &   1  \\
HR Del      &   0.76 &$ -14.0$&  0.184 &   1  \\
KT Eri      &   6.50 &$ -31.9$&  3.435 &   1  \\
DN Gem      &   0.45 &$  14.7$&  0.114 &   1  \\
DQ Her      &   0.39 &$  26.4$&  0.173 &   2  \\
V533 Her    &   0.56 &$  24.3$&  0.230 &   1  \\
CP Lac      &   1.00 &$  -0.8$&  0.014 &   1  \\
DK Lac      &   3.90 &$  -5.4$&  0.367 &   1  \\
DI Lac      &   2.25 &$  -4.9$&  0.192 &   1  \\
IM Nor      &   3.40 &$   3.0$&  0.178 &   1  \\
RS Oph      &   1.40 &$ -10.4$&  0.253 &   1  \\
V849 Oph    &   3.10 &$  13.5$&  0.724 &   1  \\
V2487 Oph   &  12.00 &$   8.0$&  1.670 &   1  \\
GK Per      &   0.48 &$ -10.1$&  0.084 &   2  \\
RR Pic      &   0.52 &$ -25.4$&  0.223 &   2  \\
CP Pup      &   1.14 &$  -0.4$&  0.008 &   2  \\
T Pyx       &   4.50 &$  10.2$&  0.797 &   1  \\
U Sco       &  12.00 &$ -60.2$& 10.413 &   1  \\
V745 Sco    &   7.80 &$ -60.2$&  6.769 &   1  \\
EU Sct      &   5.10 &$  -2.8$&  0.249 &   1  \\
FH Ser      &   0.92 &$   5.8$&  0.093 &   1  \\
V3890 Sgr   &   7.00 &$  -6.4$&  0.780 &   1  \\
LV Vul      &   0.92 &$   0.8$&  0.013 &   1  \\
NQ Vul      &   1.28 &$   1.3$&  0.029 &   1  \\
CT Ser      &   1.43 &$  47.6$&  1.056 &   1  \\
RW UMi      &   4.90 &$  33.2$&  2.683 &   1  \\
V3888 Sgr   &   2.50 &$   5.4$&  0.235 &   1  \\
PW Vul      &   1.75 &$   5.2$&  0.159 &   1  \\
QU Vul      &   1.76 &$  -6.0$&  0.184 &   1  \\
V1819 Cyg   &   7.39 &$   4.0$&  0.515 &   1  \\
V842 Cen    &   1.14 &$   3.6$&  0.072 &   1  \\
QV Vul      &   2.68 &$   7.0$&  0.327 &   1  \\
V351 Pup    &   2.53 &$  -0.7$&  0.031 &   1  \\
HY Lup      &   1.80 &$   9.0$&  0.282 &   1  \\
CP Cru      &   3.18 &$   2.2$&  0.122 &   1  \\

\enddata
\tablenotetext{a}{(1) \citet{dar12}; (2) \citet{dow00}}
\end{deluxetable}

Following \citet{bah80} for the Galactic bulge component we have:
\begin{equation}
\rho_b(r) = \rho_{b,0}~{\mathrm{exp}[-7.669(r/re)^{1/4}] \over (r/r_e)^{7/8}},
\end{equation}
where $r$ is the radial distance from the Galactic center and
$r_e=2.7$~kpc is the scale parameter for the Galactic bulge.
The constant $\rho_{b,0}$ represents the density
of bulge novae at the center of the Galaxy.

For the disk component we can write:

\begin{equation}
\rho_d(z,x) = \rho_{d,0}~exp[-(z/h_z)-(x-r_0)/h_r],
\end{equation}

\noindent
where $\rho_{d,0}$ is the density of novae at the position of the sun,
$z$ is the distance of a nova
perpendicular to the Galactic plane,
$r_0$ is the distance from the Sun to the Galactic center,
and $x=[r^2 - z^2]^{1/2}$ is the distance of a nova
from the Galactic center in the plane of the Galaxy.

The parameters
$h_r$ and $h_z$ are the scale lengths for the exponential distributions
of novae parallel and perpendicular to the Galactic plane, respectively.
\citet{bah80} adopted
$h_r=3.5$~kpc for the radial scale length of the disk and $r_0=8$~kpc
for the distance from the sun to the Galactic center. More recent studies
have suggested a slightly shorter scale length for the disk and a somewhat
larger distance to the Galactic center. Here
we adopt more recent determinations
of $h_r=3.0\pm0.2$~kpc \citep{mcm11}
and $r_0=8.33\pm0.11$~kpc \citep{cha15}.

\begin{figure}

\includegraphics[scale=0.345]{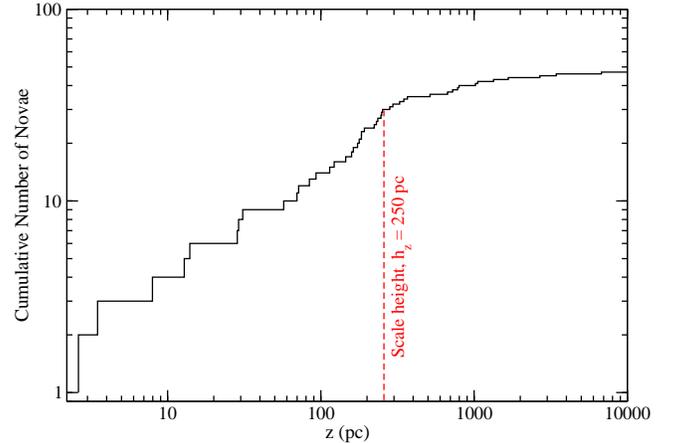} 

\caption{The $z$ distribution for Galactic novae with known distances. The
scale height in the cumulative distribution is given by the value of $z$
corresponding to $N~=~N_\mathrm{max}~(1 - e^{-1})$. Here we find $z=250$~pc.
\label{fig5}}

\end{figure}

Several estimates of the scale height of cataclysmic
variables (CVs) perpendicular to the plane of the Galaxy
have been made over the years. For example, \citet{pat84} found
$z=190$~pc for CVs in general, while \citet{due84} estimated
$z=125$~pc for Galactic novae specifically.
More recently, \citet{rev08} determined
$z=130$~pc for X-ray selected CVs, while \citet{ak08} find $z=158\pm14$~pc
for a large sample of Galactic CVs.
We have made an independent
determination of the scale height for Galactic novae using the
distance estimates given in Table~\ref{tab4}, which have been taken
from \cite{dar12} and \cite{dow00}.
Figure~\ref{fig5} shows the cumulative distribution
of scale heights for the 47 novae in this sample.
If we assume novae are distributed as $n(z)=n_0~\mathrm{exp}(-z/h_z)$, then
for a cumulative distribution where $N(z)=\int_o^zn(z)dz$, we find
that $z=h_z$ at $N = N_\mathrm{max} (1 - e^{-1})$, where
$N_\mathrm{max}=n_0h_z$. Taking $N_\mathrm{max}=47$, we find
$h_z\simeq250$~pc, which is somewhat larger than previous estimates.
In the analysis to follow, we will adopt our estimate of $h_z=250$~pc.
The adopted value of $h_z$ sets the local nova rate density,
$\rho_{d,0}$, required to normalize the model to $N(2)$.
Since the Galactic nova rate is dependent on this normalization,
unlike the value of $\rho_{d,0}$, it is not sensitive to
the choice of $h_z$.

Taken together,
$\rho_{b,0}$ and $\rho_{d,0}$ determine the relative contribution of
the bulge and disk to the overall Galactic nova rate.
At the position of the sun, \citet{bah80} find that
the bulge contributes 1/800 of the stellar density.
In this case, if we define $\theta (= n_\mathrm{disk}/n_\mathrm{bulge}$) as the ratio
of the specific nova rate of the disk population
to that of the bulge population,
we find that $\rho_{d,0} / \rho_{b,0} \simeq \theta/79$.
Assuming $\theta=1$ and a nova scale height, $h_z=250$~pc, this leads to
an integrated disk-to-bulge mass ratio of $\sim$15.
More recent models for the Galaxy suggest that the bulge component
makes up a larger fraction of the Milky Way's total
mass, and that the disk-to-bulge mass ratio, $M_d/M_b$ is
of order 6 \citep[e.g., see][]{mcm11,lic15}. Assuming the
relative nova rates follow the integrated bulge and disk masses,
a value of $M_d/M_b=6$ corresponds to
a bulge contribution of $\sim$~1/320 to the total stellar density at the
position of the sun. 
In the calculations to follow we will consider both
\citet{bah80} models characterized by $\theta=1.0$ as well as a bulge-enhanced
$\theta=0.4$ model.
The latter model is equivalent
to a $\theta=1.0$ model with a more massive bulge ($M_d/M_b = 6$),
where the bulge and disk components produce novae
at the same rate per unit mass.

\begin{figure}

\includegraphics[scale=0.345]{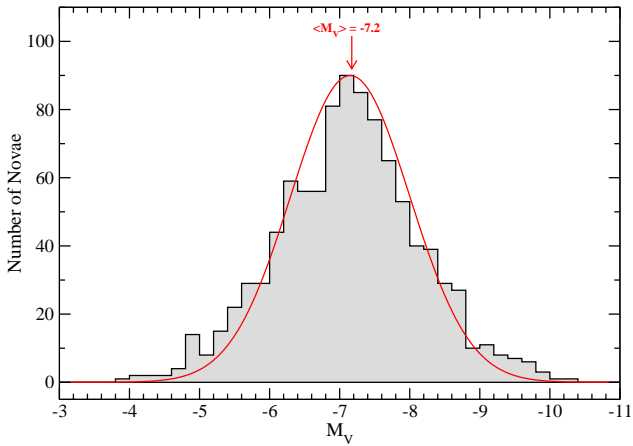} 

\caption{The absolute magnitude distribution for all known
M31 nova candidates through May 2016. The distribution is
approximately Gaussian with a mean $<$$M_V(\mathrm{M31})$$>$$~\simeq-7.2$
and standard deviation of 0.8 magnitudes. \label{fig6}}

\end{figure}

\subsection{The Absolute Magnitude Distribution of Novae}

Before we can estimate the Galactic nova rate or
compute the expected nova rate as function of apparent magnitude,
we must specify the absolute magnitude characteristic
of Galactic novae.
\citet{sha09} has shown that
the absolute magnitudes of novae typically range from $M\simeq-5$ to
$M\simeq-10$, with the mean peak absolute magnitude
for the M31 and the smaller Galaxy samles being given by
$<$$M_V(\mathrm{M31})$$>$$~\simeq-7.2$ and $<$$M_V(\mathrm{G})$$>$$~\simeq-7.8$,
respectively. To bring the M31 sample up to date,
we have redetermined absolute magnitude distribution for
the full sample of M31 novae through May 2015 given in the online catalog of
W. Pietsch\footnote{\tt see also http://www.mpe.mpg.de/$\sim$m31novae/opt/m31/index.php} using the
extinction and color corrections given in \citet{sha09}.
Figure~\ref{fig6} shows the resulting absolute magnitude distribution. The mean of the
distribution remains unchanged with $<$$M_V(\mathrm{M31})$$>$$~= -7.2$, with a
best-fitting Gaussian giving a standard deviation of 0.8~mag.

Selection effects likely bias both the Galactic and M31 nova samples.
The Galactic nova sample is affected by extinction, and
is almost certainly biased towards more luminous
(primarily disk) novae. The M31 Sample is much larger and contains novae
from both M31's bulge and disk components. However,
the maximum-light magnitude is based on discovery
magnitude rather than confirmed peak magnitude, and thus
likely underestimates the
absolute magnitude the novae at maximum light. After taking both of these
biases into account
we adopt $<$$M_V$$>$$~= -7.5\pm0.8$ as the best estimate
currently available for the average absolute magnitude distribution
of Galactic novae at maximum light.

A significant
advance in our understanding of extragalactic nova rates
came when \cite{kas11} discovered
a significant population of faint, but fast, novae in M31 that
did not appear to follow the canonical maximum-magnitude, rate-of-decline
(MMRD) relation for classical novae \citep[e.g., see][]{dow00}. A typical
example of a faint, but fast nova is the
recurrent nova M31N 2008-12a mentioned earlier,
which only reaches an absolute magnitude
$M_V~\simeq-6$ at maximum light ($m_V\simeq18.5$ at the distance of M31),
and fades by two magnitudes from peak in less than two days \citep{dar15}.
Such novae have likely been missed in the Galaxy and in previous surveys for
novae in M31. As a result of their short recurrence times, such novae
could make up a significant fraction of the observed nova rate, and,
if properly accounted for could shift the peak absolute magnitude
distribution to fainter magnitudes. To explore this possibility,
we will also take the M31 absolute magnitude distribution
at face value, and consider models where we adopt $<$$M_V$$>$$~= -7.2\pm0.8$.
Later, we will also consider the possibility that the absolute magnitudes
of bulge and disk novae differ.

\section{The Galactic Nova Rate}

\subsection{Direct Extrapolation}

\citet{sha02} estimated the global Galactic nova rate by using the
\citet{bah80} model to extrapolate the local nova rate
(assumed complete for $m\leq2$) to
sufficiently faint magnitudes that the entire galaxy
is covered.
If we define $R(m)$ as the distance from the sun to
a nova of apparent magnitude $m$ we have:

\begin{equation}
R(m) = 10^{[1 + 0.2(m - M_V - A_V(R))]}
\end{equation}

\noindent
where $M_V$ is the absolute visual magnitude at maximum light,
and $A_V(R)$ is the visual extinction suffered by a nova at distance $R$.
The value
of $A_V$ will be a function of position in the Galaxy, and is approximated
here as follows:

\begin{equation}
A_V(R) = a_V\int_{0}^{R}e^{-z/h_d}~dR = a_V R (h_d/z) (1 - e^{-z/h_d})
\end{equation}

\noindent
where the constant $a_V$ represents
the extinction in the midplane of the disk, and $h_d$ is
the scale height of the obscuring dust layer, assumed to drop
off exponentially perpendicular to the disk plane. Here, we adopt
an average extinction of $a_V=1$ mag per kpc in the Galactic midplane, with
a scale height perpendicular to the plane, $h_d=100$~pc \citep{spi78}.
Equations (4) and (5) are solved iteratively to determine $R(m)$.

The number of novae per year
visible to a given magnitude, $m$, is then given by:

\begin{equation}
N(m)=2\int_{0}^{2\pi}\int_{0}^{\pi/2}\int_{0}^{R(m)} (\rho_d+\rho_b)~R^2~dR~\mathrm{cos}~b~db~dl,
\end{equation}

\noindent
where $b$ and $l$ are the Galactic latitude and longitude, respectively, and
$\rho_b (r)$ and $\rho_d (z,x)$ from equations~(2) and (3)
can be cast in terms of $R$ by noting that

\begin{equation}
r=\big[r_0^2 + R^2 - 2 R r_0~\mathrm{cos}~b~\mathrm{cos}~l\big]^{1/2},
\end{equation}

\noindent
$z=R~\mathrm{sin}~b$, and $x=[r^2 - z^2]^{1/2}$.

\begin{figure}

\includegraphics[scale=0.345]{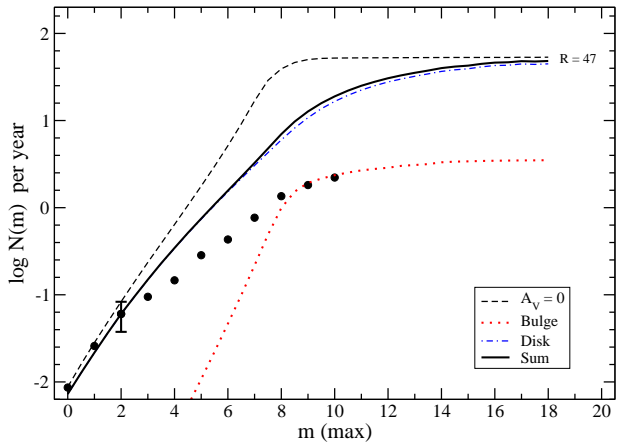} 

\caption{The cumulative nova rate distribution
as a function of peak apparent magnitude is compared
with our Galaxy model. The model has been normalized to the
observed $N(2)$ assuming 100\% completeness. The model
predicts a global nova rate of 47~yr$^{-1}$.
The dotted red line, the dot-dashed blue line,
and the solid black line represent the bulge, disk and total rate,
respectively. The dashed line represents the extinction-free
model with $A_V=0$.
\label{fig7}}

\end{figure}

\begin{figure}

\includegraphics[scale=0.345]{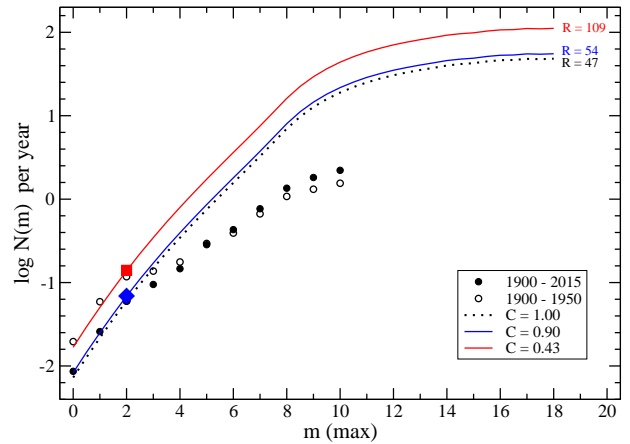} 

\caption{The same as Figure~\ref{fig7}, except that
the models have been normalized to the $c=0.9$ and
$c=0.43$ completeness-corrected $N(2)$ points shown by the blue
diamond and red square, respectively.
The dotted line shows the model assuming 100\% completeness
at $N(2)$ from Fig.~7.
Also shown as open circles are the cumulative nova rates based
on the period from 1900 to 1950. We note that the $c$=0.43 model
matches these data quite well.
\label{fig8}}

\end{figure}

Figure~\ref{fig7} shows the expected increase in the observed nova rate as a function
of apparent magnitude for the full 1900-2015 sample. The extrapolation
is accomplished by integrating equation~(4) as a function of
apparent magnitude where we assume the average nova has an
average absolute magnitude at maximum light of $<$$M_V$$>$$~= -7.5$.
The solid line shows our model for
the expected increase in the nova rate with apparent magnitude.
As in \citet{sha02}, the model has initially been normalized to the
observed value of $N(2)$ assuming
100\% completeness for $m\leq2$.
The bulge and disk contributions to
the overall nova rate are shown as the red dotted and blue dot-dashed
curves, respectively.
Given that the bulge contributes
so little to the nova density at the position of the Sun, these
values essentially reflect the nova rate densities
of disk novae. The agreement between the observed rates
and the model is quite good for novae brighter than $m=2$,
with the observed nova rates falling off with increasing
apparent magnitude as expected for a disk distribution (log $N\propto 0.4m$).
At fainter magnitudes the expected
incompleteness in the observed nova rate becomes increasingly apparent.
We note that, at magnitudes fainter than $m\simeq6$, there is a hint that the
contribution of bulge novae is beginning to kick in.

Despite the excellent fit of the model to novae with $m\leq2$,
as we have discussed in section~2, the observed nova rates
for even these bright novae are likely to be incomplete.
Figure~\ref{fig8} shows our model fits to the incompleteness-corrected
values of $N(2)$, shown by the blue diamond and the red square
for $c=0.9$ and $c=0.43$, respectively.
These models assume our best estimate
for the mean absolute magnitude of Galactic novae, $<$$M_V$$>$$~=-7.5$,
and produce
global Galactic nova rates of $\sim48$ and $\sim98$ per year
for the $c$=0.9 and $c$=0.43 models, respectively.
We note that the rate based on an assumed completeness of 43\%
for novae brighter than second magnitude
is almost exactly what we would expect if we normalized the
model to the observed rate during the period between
1900 and 1950, as shown in Figure~\ref{fig4}, and assumed 100\% completeness
during that interval.
In both Figures~\ref{fig7} and~\ref{fig8},
the overall Galactic nova rates have been approximated
as in \citet{sha02}
by extrapolating the models to sufficiently faint magnitudes that
the entire Galaxy is essentially covered. The slightly higher rates
compared with those in \citet{sha02} result from our
updated values of $h_r$ and $r_0$.

Significant limitations of the direct extrapolation
approach adopted by \citet{sha02}
and reviewed above, are that this method does not provide an uncertainty in the
derived nova rates, and that it requires that a specific
absolute magnitude for the model novae be specified
rather than allowing for a distribution of absolute magnitudes
to be considered. A roughly equivalent, but superior approach is to
extrapolate the local population of novae to the entire Galaxy
using a Monte Carlo simulation.

\subsection{Monte Carlo Simulations}

In our Monte Carlo simulations,
we distribute simulated
bulge and disk novae following the scaling laws given in equations~(2)
and~(3)
for a range of trial Galactic nova rates, $\nu$.
We then record the number of novae brighter than second magnitude,
$N_{\nu}(m\leq2)$, that are produced
over the 116 year period covered by the observations.
The magnitude of the $i^{th}$ nova is calculated
assuming the extinction given by equation~(5)
over a distance, $R_i$, from the sun given by:

\begin{equation}
R_i=\big[r_i^2 + r_0^2 - 2 r_i r_0~\mathrm{sin}~\theta_i~\mathrm{cos}~\phi_i\big]^{1/2},
\end{equation}

\noindent
where $\theta$ and $\phi$ are the usual spherical polar coordinates
with origin at the center of the Galaxy.
For the disk component, $\theta_i=\pi/2 - \mathrm{tan}^{-1}(z_i/x_i)$,
while we assume a uniform distribution in cos~$\theta$ for the bulge component.
In both cases we assume
that the Galaxy is axially symmetric (i.e., a uniform distribution in $\phi$).

\begin{figure}

\includegraphics[scale=0.4]{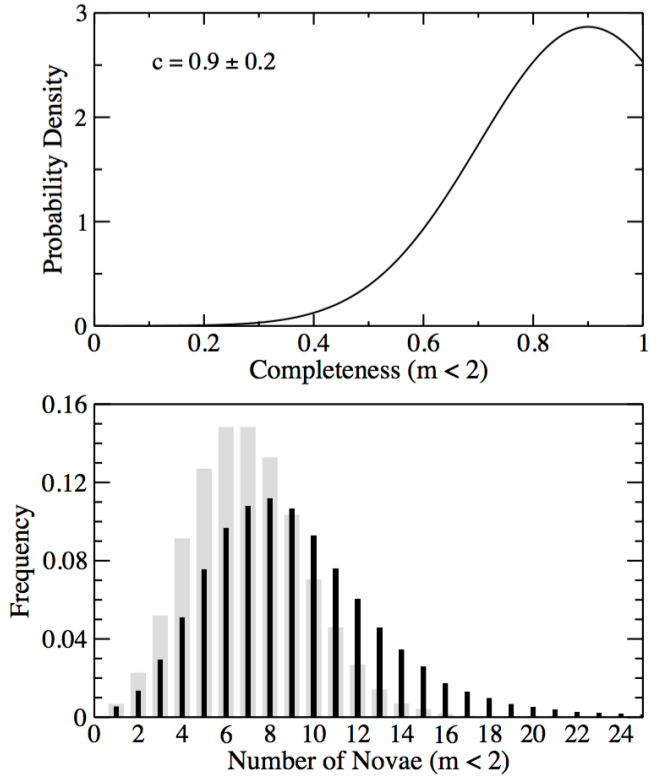} 

\caption{Top panel: The adopted completeness function for novae with $m\leq2$
given by $c=0.9\pm0.2$.
Bottom panel: The frequency distribution of $N_c(m\leq2)$ used in the
Monte Carlo simulation (black bars), which is based on a Poisson distribution
(mean of 7) for $N\mathrm{obs}$ (grey bars)
convolved with the completeness function shown in the top panel.
\label{fig9}}

\end{figure}

We run the simulation
$M=10^5$ times for each trial nova rate
and record the number of times $N_{\nu,i}(m\leq2)$ from the
$i^{th}$ simulation matches
the number of novae believed to have
reached second magnitude or brighter over the past 116 years.
This latter number is
simply given by $N_c(m\leq2) = [N_\mathrm{obs}/c]$, where
$c$ is the assumed observational completeness for $m\leq2$.
To account for variation in the number of $m\leq2$ novae observed
over the past 116 years, our Monte Carlo simulations
sample a Poisson distribution (mean of 7) for $N_\mathrm{obs}$.
Similarly,
to account for the uncertainty in $c$, our Monte Carlo
simulation samples a distribution, $c_i$,
which is normally distributed about $c$ with standard deviation $\sigma_c$.
Figure~\ref{fig9} shows the completeness function and the resulting
$N_{c,i}(m\leq2)$ distribution
for our $c=0.9$ models, where we have assumed $\sigma_c=0.2$.
The most likely estimate of the global nova rate for a given
assumed completeness distribution is then given
by the value of $\nu$ that produces the largest number of matches
between $N_{\nu,i}(m\leq2)$ and $N_{c,i}(m\leq2)$. Specifically,
the probability of a given nova rate is given by:

\begin{equation}
P(\nu) = {\sum_{i=1}^{M} \delta_{N_{\nu,i}, N_{c,i}} \over \sum_{\nu=0}^{\infty} \sum_{i=1}^{M} \delta_{N_{\nu,i}, N_{c,i}}},
\end{equation}

\noindent
where $\delta$ is the Kronecker delta function.

We have run an array of Monte Carlo simulations to explore
how the choice of model parameters affect the Galactic nova rate.
We initially assumed that
the nova rate
per unit mass is the same in the disk and bulge (i.e., $\theta=1$),
and considered both a disk-to-bulge mass ratio
$M_d/M_b = 15$ as found by \citet{bah80} in their pioneering
study of the Milky way, and
a bulge-enhanced model with $M_d/M_b = 6$.
The latter model is equivalent a \citet{bah80} model with a
higher specific bulge nova rate given by $\theta=0.4$,
which is the value
found by \citet{sha01} based on the spatial distribution of novae in M31.
For comparison,
we have also computed a pure disk model ($\theta=\infty$) despite the fact that
available observations do not support a heavily disk-dominated nova
population.

\begin{figure}

\includegraphics[scale=0.345]{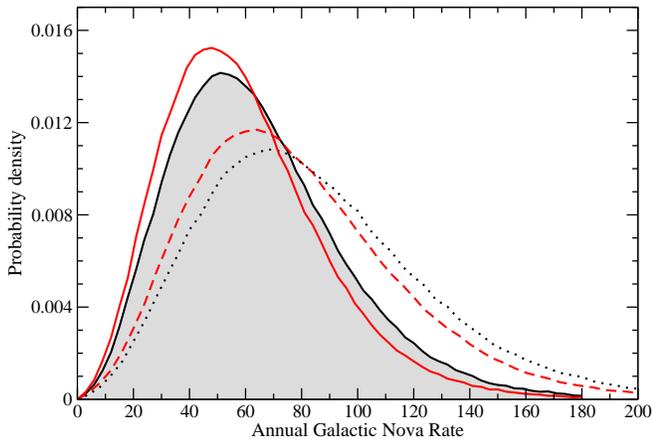} 

\caption{Galactic nova rate probability density
distributions from the Monte Carlo simulations.
The shaded region bounded by the black solid line represents our preferred
model characterized by $<$$M_V$$>$$~= -7.5\pm0.8$ and $\theta=0.4$.
The red solid curve shows the distribution for a smaller bulge fraction
characterized by $\theta=1.0$. The broken lines represent nova
rate probabilities for $<$$M_V$$>$$~= -7.2\pm0.8$, with the dashed line
for $\theta=1$, and the dotted line for $\theta=0.4$.
The peak of a distribution represents the most probable nova rate
for a given model. \label{fig10}}

\end{figure}

Figure~\ref{fig10} shows a plot of $P(\nu)$
as a function of the trial
global nova rate $\nu$
for our most plausible models.
Four $c=0.9\pm0.2$ models are shown
representing differences both in
the contribution of the Milky Way's bulge and disk components and
in the assumed absolute magnitude distribution of novae.
The solid red curve shows the nova rate
distribution for a Galactic nova population characterized
by $<$$M_V$$>$$~= -7.5\pm0.8$ and
an assumed disk-to-bulge ratio of 15 ($\theta=1$), while
the shaded distribution represents our preferred model with
an enhanced bulge component ($\theta=0.4$)
that contributes $1/6$ that of the Milky Way's disk
\citep[e.g., see][]{lic15,mcm11}.
The peak of this distribution corresponds to a most likely
nova rate of $52^{-23}_{+32}$ novae per year, where the uncertainties
are 1$\sigma$ errors of an assumed bi-Gaussian distribution \citep{buy72}.
Since the local novae ($m\lessim2$) are predominately disk novae, an increase
in the bulge-to-disk ratio has the effect of shifting the peak of the
probability distribution to somewhat higher nova rates.

The broken lines in Figure~\ref{fig10}
show nova rate distributions for the two disk-to-bulge
ratios assuming the fainter
absolute magnitude distribution observed for M31, $<$$M_V$$>$$~= -7.2\pm0.8$
(see Fig.~5). These models may be more appropriate for our Galaxy
if a significant
population of faint and fast novae have escaped detection
in previous extragalactic nova surveys \citep[e.g.,][]{kas11}. If so,
$<$$M_V$$>$$~= -7.5\pm0.8$
may overestimate the peak luminosity of the Galactic nova
distribution.
As expected, a lower assumed mean nova
luminosity reduces the volume sampled by $m\leq2$ novae, increases the
extrapolation to the entire Galaxy, and results in
overall Galactic nova rates as high as $\sim65$ per year.

In addition to our preferred $c=0.9\pm0.2$ models,
we have also computed model rates based on the low completeness
of $c=0.43\pm0.06$ suggested by \citet{she14}. As expected, these
models result in considerably higher nova
rates that reach of order 100 per year in the case of the
$<$$M_V$$>$$~= -7.5\pm0.8$ models, or even higher
if we adopt the lower mean luminosity of
$<$$M_V$$>$$~= -7.2\pm0.8$.
As discussed earlier, we consider it
extremely unlikely that the completeness of second magnitude
and brighter novae could be below 50\%. Thus, the nova rates
that result from the $c=0.43\pm0.06$ models likely represent
firm upper limits to the Galactic nova rate.

\begin{deluxetable*}{lccccccccc}
\tablenum{5}
\label{tab5}
\tablewidth{0pt}
\tablecolumns{10}
\tablecaption{Galactic Nova Rate Estimates}
\tablehead{\colhead{} & \colhead{} & \colhead{} & \colhead{} & \colhead{$c=0.9\pm0.2$} & \colhead{} & \colhead{} & \colhead{} & \colhead{$c=0.43\pm0.06$} & \colhead{} \\
\colhead{} & \colhead{} & \colhead{} & \colhead{Bulge} & \colhead{Disk} & \colhead{Total} & \colhead{} & \colhead{Bulge} & \colhead{Disk} & \colhead{Total} \\
\colhead{($\theta$)} & \colhead{($M_d/M_b$)} & \colhead{($\rho_d/\rho_b)_{\odot}$} & \colhead{Rate (yr$^{-1}$)} & \colhead{Rate (yr$^{-1}$)} & \colhead{Rate (yr$^{-1}$)} & \colhead{} & \colhead{Rate (yr$^{-1}$)} & \colhead{Rate (yr$^{-1}$)} & \colhead{Rate (yr$^{-1}$)} }

\startdata

\cutinhead{$<$$M_{V,\mathrm{disk}}$$>$$~= -7.5\pm0.8,~$$<$$M_{V,\mathrm{bulge}}$$>$$~= -7.5\pm0.8$}
 1.0       &  15      & 800    & $3.0_{-1.4}^{+1.9}$   & $45_{-21}^{+28}$ & $48_{-22}^{+30}$ & & $5.7_{-1.9}^{+2.5}$  & $84_{-28}^{+37}$  & $90_{-30}^{+40}$\\
 0.4 (1.0) &  15 (6)  & 320    & $7.4_{-3.4}^{+4.6}$   & $45_{-21}^{+27}$ & $52_{-24}^{+32}$ & & $14_{-4.6}^{+6.3}$   & $84_{-27}^{+38}$  & $98_{-32}^{+44}$\\
 $\infty$  & $\infty$ & $\infty$ & \dots               & $45_{-19}^{+27}$ & $45_{-19}^{+27}$ & & \dots                & $86_{-30}^{+37}$  & $86_{-30}^{+37}$\\
\cutinhead{$<$$M_{V,\mathrm{disk}}$$>$$~= -7.2\pm0.8,~$$<$$M_{V,\mathrm{bulge}}$$>$$~= -7.2\pm0.8$}
 1.0       &  15      & 800    & $4.0_{-1.8}^{+2.3}$   & $60_{-27}^{+35}$ & $64_{-29}^{+37}$ & & $7.6_{-2.5}^{+3.2}$  & $112_{-37}^{+48}$ & $120_{-40}^{+51}$\\
 0.4 (1.0) &  15 (6)  & 320    & $9.7_{-4.3}^{+6.0}$    & $58_{-26}^{+36}$ & $68_{-30}^{+42}$ & & $19_{-6.3}^{+8.0}$   & $111_{-38}^{+48}$ & $130_{-44}^{+56}$\\
 $\infty$  & $\infty$ &$\infty$& \dots                 & $60_{-27}^{+37}$ & $60_{-27}^{+37}$ & & \dots                & $112_{-37}^{+49}$ & $112_{-37}^{+49}$\\
\cutinhead{$<$$M_{V,\mathrm{disk}}$$>$$~= -7.8\pm0.8,~$$<$$M_{V,\mathrm{bulge}}$$>$$~= -7.2\pm0.8$}
 1.0       &  15      & 800    & $2.3_{-1.2}^{+1.8}$  & $34_{-18}^{+26}$ & $36_{-19}^{+28}$ & & $4.3_{-1.4}^{+1.9}$ & $65_{-21}^{+29}$  & $69_{-23}^{+31}$\\
 0.4 (1.0) &  15 (6)  & 320    & $5.7_{-2.6}^{+3.6}$   & $34_{-15}^{+21}$  & $40_{-18}^{+25}$ & & $11_{-3.6}^{+4.7}$  & $64_{-21}^{+28}$  & $75_{-25}^{+33}$\\
 $\infty$  & $\infty$ &$\infty$& \dots                 & $35_{-16}^{+21}$  & $35_{-16}^{+21}$ & & \dots               & $65_{-23}^{+30}$  & $65_{-23}^{+30}$\\

\enddata

\end{deluxetable*}

A summary of our full array of models is given in Table~\ref{tab5}.
Each model is described in columns $1-3$, which give the
ratio of the specific disk-to-bulge nova rates, $\theta$,
the disk-to-bulge mass ratio, $M_d/M_b$, and the corresponding
disk-to-bulge nova density ratio at the position of the sun,
($\rho_d/\rho_b)_{\odot}$.
The derived nova rates are not sensitive to the adopted model
parameters such as the bulge scale parameter, $r_e$, or the disk
scale length, $h_r$, and height, $h_z$.
The local nova rate density $\rho_{d,0}$, however, does
depend on the assumed scale height of novae in the disk.
Values of $\rho_{d,0}$
for three representative values of $h_z$ are given in Table~\ref{tab6}.

\begin{deluxetable}{cccc}
\tablenum{6}
\label{tab6}
\tablewidth{0pt}
\tablecolumns{4}
\tablecaption{Local Nova Rate Density Estimates}
\tablehead{\colhead{} & \colhead{$c=1.0$} & \colhead{$c=0.9\pm0.2$} & \colhead{$c=0.43\pm0.06$}\\
\colhead{$h_z$} & \colhead{$\rho_{d,0}$}& \colhead{$\rho_{d,0}$}  & \colhead{$\rho_{d,0}$}\\
\colhead{(pc)} & \colhead{(kpc$^{-3}$~yr$^{-1}$)}& \colhead{(kpc$^{-3}$~yr$^{-1}$)}  & \colhead{(kpc$^{-3}$~yr$^{-1}$)}  }

\startdata

\cutinhead{$<$$M_{V,\mathrm{disk}}$$>$$~= -7.5,~$$<$$M_{V,\mathrm{bulge}}$$>$$~= -7.5$}
\\
 125& 0.18&0.21 & 0.39    \\
 150& 0.16&0.19 & 0.34   \\
 250& 0.11&0.13 & 0.24    \\
\cutinhead{$<$$M_{V,\mathrm{disk}}$$>$$~= -7.2,~$$<$$M_{V,\mathrm{bulge}}$$>$$~= -7.2$}
\\
 125& 0.23&0.26 & 0.50    \\
 150& 0.21&0.25 & 0.49   \\
 250& 0.15&0.17 & 0.32    \\
\cutinhead{$<$$M_{V,\mathrm{disk}}$$>$$~= -7.8,~$$<$$M_{V,\mathrm{bulge}}$$>$$~= -7.2$}
\\
 125 & 0.12  &0.13  & 0.28    \\
 150 & 0.11  &0.12  & 0.25    \\
 250 & 0.074 &0.082 & 0.18    \\

\enddata

\end{deluxetable}

We note that Galactic nova rates determined
in our Monte Carlo simulations are
slightly lower than those found in the direct extrapolation
described in the previous section.
This difference is the result of the fact that for a given apparent
magnitude, the volume, $V(M_V)$,
sampled is not directly proportional to $M_V$,
but rather $V(M_V)~\propto 10^{-0.6M_V}$.
Thus, the {\it average\/} volume sampled for a Gaussian distribution
of absolute magnitudes is larger than the volume sampled assuming
a single average absolute magnitude alone.
In other words, $<$$V(M_V)$$>$~$>$~$V($$<$$M_V$$>$). The larger volume
sampled for novae with $m\leq2$ results in a smaller extrapolation,
and therefore a lower overall nova rate.

\subsubsection{The Effect of Differing Nova Populations}

For some time there has been speculation that
that there may be two distinct populations of novae:
a bulge population characterized by relatively slow and dim
novae, and a disk population characterized by somewhat brighter novae
that lie closer to the Galactic plane and generally evolve
more quickly \citep{del92}.
The evidence for two populations finds some support from
observations showing that novae can be divided into two classes
based upon the character of their spectra shortly after eruption.
Although all novae display prominent Balmer emission shortly after eruption,
the so-called ``Fe~II novae", are characterized by relatively narrow
($\sim$1000~km~s$^{-1}$ FWHM) Fe~II emission
features that exhibit P-Cyg profiles near maximum light, while the
``He/N novae" show prominent and broad ($\grtsim$2500~km~s$^{-1}$ FWHM)
He~I and N emission features, but without the Fe~II emission
\cite{wil92}.
The work of \cite{del98} showed that the
He/N novae cluster close to the Galactic plane and tend to be fast and
bright relative to the Fe~II class.

Given the possible existence of two classes of novae, we have also
computed nova rate models that are characterized by
different average peak luminosities and specific nova rates
in the Galactic bulge and disk. Specifically, we have considered
a two-component Galaxy model where disk novae are
characterized by $<$$M_V$$>$$~=-7.8$ and bulge novae by
$<$$M_V$$>$$~=-7.2$. Since the
$m\leq2$ disk novae used in the normalization are more luminous,
they extend to a larger volume of the Galaxy.
Thus, both the density of
novae in the vicinity of the sun ($\rho_{d,0}$)
and the extrapolated global Galactic
nova rate are significantly reduced.
Bulge-dominated ($\theta=0.4$) models
produce generally higher rates because the normalization is set by
nearby novae that belong overwhelmingly to the disk, whereas
disk dominated models ($\theta=\infty$) produce lower rates because there
is no contribution from the bulge.

\begin{deluxetable}{cccccc}
\tablenum{7}
\label{tab7}
\tablewidth{0pt}
\tablecolumns{6}
\tablecaption{Predicted Visibility of Galactic Novae}
\tablehead{\colhead{} & \multicolumn2c{$c=0.9\pm0.2$} & \colhead{} & \multicolumn2c{$c=0.43\pm0.06$} \\
\colhead{} & \colhead{$\theta=1.0$} & \colhead{$\theta=0.4$} & \colhead{} & \colhead{$\theta=1.0$} & \colhead{$\theta=0.4$} \\
\colhead{$m_{\mathrm lim}$} & \colhead{$N$~(yr$^{-1})$} & \colhead{$N$~(yr$^{-1})$} & \colhead{} & \colhead{$N$~(yr$^{-1})$} & \colhead{$N$~(yr$^{-1})$} }

\startdata
\cutinhead{$<$$M_{V,\mathrm{disk}}$$>$$~= -7.5,~$$<$$M_{V,\mathrm{bulge}}$$>$$~= -7.5$}
\\
  2.0    & 0.1 &  0.1    &&    0.1  &    0.1 \\
  4.0    & 0.4 &  0.4    &&    0.7  &    0.7 \\
  6.0    & 1.5 &    1.5  &&    2.8  &    2.8 \\
  8.0    & 5.2 &    5.4  &&    9.7  &   10   \\
 10.0    &14   &   14    &&   25    &   27   \\
 12.0    &23   &   25    &&   44    &   48   \\
 14.0    &32   &   35    &&   60    &   66   \\
 16.0    &38   &   42    &&   72    &   79   \\
 18.0    &42   &   46    &&   80    &   87   \\
 20.0    &45   &   49    &&   84    &   92   \\

\cutinhead{$<$$M_{V,\mathrm{disk}}$$>$$~= -7.2,~$$<$$M_{V,\mathrm{bulge}}$$>$$~= -7.2$}
\\
  2.0    & 0.1 &  0.1    &&    0.1  &    0.1 \\
  4.0    & 0.4 &  0.4    &&    0.7  &    0.7 \\
  6.0    & 1.6 &    1.6  &&    3.0  &    3.0 \\
  8.0    & 5.8 &    5.8  &&   11    &   11   \\
 10.0    &16   &   17    &&   30    &   32   \\
 12.0    &29   &   31    &&   55    &   59   \\
 14.0    &41   &   44    &&   77    &   84   \\
 16.0    &50   &   54    &&   94    &  102   \\
 18.0    &56   &   60    &&  105    &  114   \\
 20.0    &59   &   63    &&  111    &  121   \\

\cutinhead{$<$$M_{V,\mathrm{disk}}$$>$$~= -7.8,~$$<$$M_{V,\mathrm{bulge}}$$>$$~= -7.2$}
\\
  2.0    & 0.1 &  0.1    &&    0.1  &    0.1 \\
  4.0    & 0.3 &  0.3    &&    0.7  &    0.7 \\
  6.0    & 1.3 &    1.4  &&    2.6  &    2.6 \\
  8.0    & 4.5 &    4.7  &&    8.7  &    8.9 \\
 10.0    & 11  &   12    &&   21    &   22   \\
 12.0    & 19  &   20    &&   35    &   38   \\
 14.0    & 25  &   27    &&   47    &   51   \\
 16.0    & 29  &   33    &&   56    &   61   \\
 18.0    & 32  &   36    &&   62    &   67   \\
 20.0    & 34  &   38    &&   65    &   70   \\

\enddata
\end{deluxetable}

\subsubsection{Bulge Rates}

In addition to summarizing Galactic nova rates produced by the
various models, Table~\ref{tab5} also breaks down the
corresponding bulge (and disk) contributions to the overall rate.
As with the overall rates,
the bulge nova rates also depend on the adopted Galaxy model,
absolute magnitude distribution,
and completeness at bright magnitudes, $c$.
Our $c=0.9\pm0.2$ models predict Galactic bulge rates ranging between
$\sim3$~yr$^{-1}$ and $\sim10$~yr$^{-1}$ depending on assumed parameters.
The lowest bulge rates result from our
$\theta=1$, $<$$M_{V}$$>$$~= -7.5$ models, with the
$\theta=0.4$, $<$$M_{V}$$>$$~= -7.2$ 
variations producing significantly higher rates.
As expected, the $c=0.43\pm0.06$ models
produce bulge rates that are approximately a factor of two higher.
The bulge rate of $13.8\pm2.6$ found by \citet{mro15},
would appear most consistent with either the bulge-enhanced ($\theta=0.4$),
$c=0.9\pm0.2$ models or the $\theta=1$, $c=0.43\pm0.06$ models,
which produce bulge
nova rates of $\sim7-10$~yr$^{-1}$ and $15-20$~yr$^{-1}$, respectively.

\subsubsection{The Predicted Nova Rate as a Function of Apparent Magnitude}

In addition to estimating the overall Galactic nova rate,
our Monte Carlo simulations have been used to predict
how the number of Galactic novae should increase with
apparent magnitude. Assuming the most probable overall nova rates
from Table~\ref{tab5}, we show in
Table~\ref{tab7} the predicted increase in the number of novae
visible as a function of apparent magnitude for our
$\theta=1.0$ and $\theta=0.4$ models with and
assumed completenesses of $c=0.9\pm0.2$ and $c=0.43\pm0.06$.
These predictions can be compared with the results from
several ongoing and planned all-sky surveys such as ASAS-SN\footnote{http://www.astronomy.ohio-state.edu/$\sim$assassin/index.shtml} and ZTF\footnote{http://www.ptf.caltech.edu/ztf} once they are available, and used to differentiate
between the various Galactic nova rate models presented here.

\section{Comparison with Extragalactic Nova Rates}

In recent years evidence has been building that extragalactic
nova rates, which have been determined primarily through
synoptic surveys often with sparse temporal sampling,
may have been systematically underestimated. In particular,
\citet{cur15} and \citet{sha16} have recently argued that the nova rate
in the giant Virgo elliptical galaxy M87 is likely $2-3$ times larger
than previously thought, with the latter authors suggesting that the
$K$-band luminosity-specific nova rates (LSNRs)
of all galaxies may be $\sim3-4$ times
higher than the previous average of $\sim$2 novae per year per
$10^{10}$ solar luminosities in $K$ \citep{sha14}.

As mentioned earlier, the work of \citet{kas11} has shown that
a population of faint and fast novae, which deviate from the
canonical MMRD relation may exist. Owing to their intrinsically
low peak luminosities and their rapid declines,
such novae have almost certainly been missed in previous
magnitude-limited and low cadence synoptic surveys for novae in M31.
Because these novae deviate so strongly from the assumed MMRD, they
have not been properly accounted for when the surveys were corrected for
completeness. In most surveys, the incompleteness, and thus the final nova
rates, have likely been underestimated.

The generally accepted nova rate for M31 is $65^{+16}_{-15}$ \citep{dar06}.
This value has been called into question
recently by \citet{che16} who have computed population synthesis models
to estimate nova rates in galaxies with differing star formation histories
and morphological types. They estimate a global nova rate for
M31 of 97~yr$^{-1}$. In another study, \citet{sor16} corrected the nova samples
of \citet{arp56} and \citet{dar06} for bias against the discovery
of fast novae in these synoptic surveys. Based on these corrections,
they estimate a global nova rate for
M31 of order 106~yr$^{-1}$. Thus, the most recent estimates suggest
the nova rate in M31 could be as high as $\sim100$~yr$^{-1}$.
The stellar mass of M31 relative to the Galaxy
is not precisely known, but we can compare estimates for the integrated
$K$-band luminosity, which should reflect the mass difference.
For the Galaxy, we adopt $M_V=-20.6$ \citep{bah80}, while
for M31 we have $m_V=3.44$ \citep{dev91} and $(m-M)_0=24.38$ \citep{fre01},
yielding $M_V(\mathrm{M31})=-20.94$.
Since both galaxies have similar morphological types (Sbc), their integrated
$V-K$ colors should be similar, and in both cases we adopt
$<$$V-K$$>~\simeq3.25$ \citep{aar78}. Thus, we estimate
$M_K(\mathrm{M31})=-24.19$ and $M_K(\mathrm{MW})=-23.85$, which corresponds to
a luminosity ratio, $L_{K,\mathrm{MW}}/L_{K,\mathrm{M31}}\simeq0.73$.
Assuming the relative nova rates follow the relative luminosities,
based on a comparison with M31,
we expect a nova rate in the Galaxy of between $\sim$50~yr$^{-1}$
and $\sim70$~yr$^{-1}$, depending on whether we adopt the \citet{dar06}
or the recent estimates. Generally speaking, these estimates are in good
agreement with an average of the rates given in Table~\ref{tab5}.
The corresponding LSNRs for both
M31 and the Galaxy are $\sim8.5\pm1.5$
novae per year per $10^{10}$~$L_{\odot,K}$.
This value is in excellent agreement with LSNRs of
$9.2^{+2.7}_{-3.0}$ and $8.6\pm2.7$
novae per year per $10^{10}$~$L_{\odot,K}$ recently found by \citet{sha16} and
\citet{mro16} for M87 and the Galactic bulge, respectively.

\section{Conclusions}

By considering how the observed nova rate in the Galaxy varies with
apparent magnitude,
\citet{sha02} estimated the overall Galactic nova rate by extrapolating
the observed rate in the vicinity of the sun ($m\lessim2$), which
was assumed to be complete, to the entire
Galaxy using a two-component disk-plus-bulge model similar to that
employed by \citet{bah80} to model the stellar density in the Milky Way.
In this paper, we have reconsidered and improved the \citet{sha02} analysis
in two important respects. First, we have considered important corrections
for the incompleteness in the observed sample of nearby
novae with $m\leq2$, and secondly,
we have employed a Monte Carlo analysis to extrapolate the
local nova rate to the entire galaxy. The Monte Carlo analysis
has the advantage of allowing us to consider a distribution
of nova absolute magnitudes rather than a single mean value as in
\citet{sha02} and to better assess the uncertainties in our
derived Galactic nova rate estimates.
In addition to these two principal improvements, we
have also updated the sample of novae that have become available over the past
15 years (albeit with no new novae discovered with $m\leq2$),
and updated the values for several Galaxy model parameters.

Galactic nova rate estimates based upon our array of models
have been summarized in Table~\ref{tab5}.
We have considered disk to bulge mass ratios, $M_d/M_b=15$,
as found by \citet{bah80}, as well as a more massive bulge model
characterized by $M_d/M_b=6$. The latter model is equivalent to
a \citet{bah80} model where the specific nova rate in the disk
is just 0.4 times that of the bulge ($\theta=1.0$ vs $\theta=0.4$).
Our models have been normalized
assuming correction factors of $c=0.9\pm0.2$ and $c=0.43\pm0.06$
for the observed fraction
of novae that reached second magnitude or brighter since 1900.
The $c=0.9\pm0.2$ models represent our best estimate of the actual completeness
of novae with $m\leq2$ observed since 1900, with the $c=0.43\pm0.06$ models
representing a likely lower limit to the completeness, and thus
an upper limit to the nova rate.
We have also considered models assuming
two different nova absolute magnitude distributions. Based on both
Galactic and M31 nova samples, we consider 
$<$$M_V$$>$$~=-7.5\pm0.8$ to represent the best estimate available
for the absolute magnitude distribution of Galactic novae. For comparison,
we also considered a somewhat fainter absolute magnitude distribution
given by $<$$M_V$$>$$~=-7.2\pm0.8$, which is characteristic
of the observed distribution in M31. The latter absolute magnitude
distribution may be more appropriate if there is a significant population
of faint and ``fast" novae that have hitherto largely escaped detection
in existing surveys. Our models produce Galactic nova rates that typically
range from $\sim$50 per year to in excess of 100 per year
in the case where we are observing only 43\% of the novae
reaching $m=2$ or brighter. We have also considered models where the
absolute magnitudes of the disk and bulge nova rates differ.
If disk novae are significantly more
luminous than bulge novae (e.g., $<$$M_{V,\mathrm{disk}}$$>$$=-7.8\pm0.8$ and
$<$$M_{V,\mathrm{bulge}}$$>$$=-7.2\pm0.8$),
then somewhat lower nova rates result, falling in the range
of $\sim35$ to $\sim$75 per year.

Since we currently
know little about how nova properties may vary with population,
and the completeness estimates are similarly uncertain,
arriving at a definitive estimate for the Galactic nova rate based
on existing data is not possible.
Taking an average of the most plausible $c=0.9\pm0.2$ rates from Table~\ref{tab5}
(the $\theta=1$ and $\theta=0.4$ models),
yields a value of $50_{-23}^{+31}$~yr$^{-1}$,
which we adopt as the best estimate
currently available for the nova rate in the Galaxy.
Rates on the order of 100 per year are possible in the event that
we have missed roughly half of the novae that have reached second
magnitude or brighter over the last century.
Despite the large uncertainties, the rates derived in the present study
point towards significant increases over recent estimates,
and bring the Galactic nova rate into better agreement with that expected
based on comparison with the latest results from extragalactic surveys.

\acknowledgments
I thank an anonymous referee for valuable suggestions that
resulted in a significant improvement in the analysis.
I have also benefited from fruitful discussions with C. T. Daub, M. Henze,
and B. E. Schaefer who first recognized that the historical
record of the brightest novae was likely incomplete.
C. T. Daub, M. Henze and P. Mr\'oz
also provided valuable comments on an earlier version of the manuscript.
Financial support through NSF grant AST1009566 is gratefully acknowledged.

\end{document}


\begin{deluxetable}{lllrrrr}
\tabletypesize{\scriptsize}
\tablenum{1}
\label{tab1}
\tablewidth{0pt}
\tablecolumns{7}
\tablecaption{Galactic Novae Since 1900}
\tablehead{\colhead{Nova} & \colhead{Name} & \colhead{Date (Y,M,D)} & \colhead{$l$ (deg)} & \colhead{$b$ (deg)} & \colhead{$m_{dis}$\tablenotemark{a}} &  \colhead{Ref\tablenotemark{b}} }

\startdata

N Sgr 2015\#3   & V5669 Sgr  & 2015   9  27 & $   3.0$ & $ -2.8$ &  8.7 & 1 \\
N Sgr 2015\#2   & V5668 Sgr  & 2015   3  15 & $   7.1$ & $ -9.1$ &  4.0 & 1 \\
N Sgr 2015\#1   & V5667 Sgr  & 2015   2  12 & $   7.4$ & $ -3.2$ &  9.0 & 1 \\
N Sco 2015\#1   & V1535 Sco  & 2015   2  11 & $ 350.0$ & $  4.0$ &  8.2 & 1 \\
RN Sco 1937    & V745 Sco   & 2014   2   6 & $ 357.8$ & $ -3.8$ &  9.0 & 1 \\
N Sgr 2014     & V5666 Sgr  & 2014   1  26 & $  11.0$ & $ -4.1$ &  8.7 & 1 \\
N Cen 2013     & V1369 Cen  & 2013  12   2 & $ 311.1$ & $  3.0$ &  3.3 & 1 \\
N Del 2013     & V339 Del   & 2013   8  14 & $  62.2$ & $ -9.4$ &  4.3 & 1 \\
N Cep 2013     & V809 Cep   & 2013   2   2 & $ 110.6$ & $  0.4$ &  9.8 & 1 \\
N Mon 2012     & V959 Mon   & 2012   8   9 & $ 206.3$ & $  0.1$ &  9.4 & 1 \\
N Sgr 2012\#4   & V5592 Sgr  & 2012   7   7 & $   6.1$ & $ -5.4$ &  7.7 & 1 \\
N Sgr 2012\#3   & V5591 Sgr  & 2012   6  26 & $   8.0$ & $  3.0$ &  9.0 & 1 \\
N Sco 2012     & V1324 Sco  & 2012   5  22 & $ 358.5$ & $ -2.2$ &  9.0 & 1 \\
N Oph 2012\#2   & V2677 Oph  & 2012   5  19 & $   4.2$ & $  4.1$ &  9.5 & 1 \\
N Sgr 2012     & V5589 Sgr  & 2012   4  21 & $   5.1$ & $  3.2$ &  9.0 & 1 \\
N Cen 2012     & V1368 Cen  & 2012   3  23 & $ 309.5$ & $  4.5$ &  8.4 & 1 \\
N Sco 2011\#2   & V1313 Sco  & 2011   9   6 & $ 342.4$ & $  4.6$ &  9.1 & 1 \\
N Lup 2011     & PR Lup     & 2011   8   4 & $ 320.0$ & $  3.8$ &  8.5 & 1 \\
N Sco 2011     & V1312 Sco  & 2011   6   1 & $ 347.1$ & $  3.8$ &  9.5 & 1 \\
RN Pyx 1890    & T Pyx      & 2011   4  14 & $ 256.6$ & $ 10.2$ &  6.0 & 1 \\
N Sco 2010\#2   & V1311 Sco  & 2010   4  25 & $ 346.6$ & $  3.5$ &  8.6 & 1 \\
N Oph 2010\#2   & V2674 Oph  & 2010   2  18 & $ 359.2$ & $  4.5$ &  9.4 & 1 \\
RN Sco 1863    & U Sco      & 2010   1  28 & $ 359.1$ & $ 23.0$ &  7.5 & 1 \\
N Sgr 2010\#1   & V5585 Sgr  & 2010   1  20 & $   2.4$ & $ -4.2$ &  8.5 & 1 \\
N Oph 2010\#1   & V2673 Oph  & 2010   1  15 & $   6.6$ & $  5.7$ &  8.8 & 1 \\
N Eri 2009     & KT Eri     & 2009  11  25 & $ 207.6$ & $-31.9$ &  5.4 & 1 \\
N Sgr 2009\#1   & V5584 Sgr  & 2009  10  26 & $  16.7$ & $ -2.8$ &  9.3 & 1 \\
N Sct 2009     & V496  Sct  & 2009  10   8 & $  26.4$ & $ -1.2$ &  8.8 & 1 \\
N Oph 2009     & V2672 Oph  & 2009   8  16 & $   2.3$ & $  3.3$ & 10.0 & 1 \\
N Sgr 2009\#3   & V5583 Sgr  & 2009   8   6 & $ 359.5$ & $ -5.7$ &  7.7 & 1 \\
N Cen 2009     & V1213 Cen  & 2009   5   8 & $ 307.6$ & $  0.5$ &  8.5 & 1 \\
N Sgr 2008\#2   & V5580 Sgr  & 2008  11  29 & $   4.8$ & $ -6.5$ &  8.0 & 1 \\
N Car 2008     & V679 Car   & 2008  11  26 & $ 291.3$ & $ -0.1$ &  7.6 & 1 \\
N Mus 2008     & QY Mus     & 2008   9  28 & $ 305.5$ & $ -3.6$ &  8.6 & 1 \\
N Sco 2008     & V1309 Sco  & 2008   9   2 & $   1.0$ & $ -2.4$ &  9.5 & 1 \\
N Cen 2008     & V1212 Cen  & 2008   8  26 & $ 314.0$ & $ -3.3$ &  8.4 & 1 \\
N Sgr 2008     & V5579 Sgr  & 2008   4  18 & $   4.1$ & $ -2.8$ &  6.7 & 1 \\
N Cyg 2008\#2   & V2491 Cyg  & 2008   4  10 & $  67.2$ & $  4.4$ &  7.1 & 1 \\
N TrA 2008     & NR TrA     & 2008   4   1 & $ 326.6$ & $ -6.6$ &  9.0 & 1 \\
N Cyg 2008\#1   & V2468 Cyg  & 2008   3   7 & $  66.8$ & $  0.2$ &  7.4 & 1 \\
N Vul 2007\#2   & V459 Vul   & 2007  12  25 & $  58.2$ & $ -2.2$ &  7.7 & 2 \\
N Pup 2007\#1   & V597 Pup   & 2007   8  14 & $ 252.1$ & $  0.8$ &  7.0 & 2 \\
N Vul 2007\#1   & V458 Vul   & 2007   8   4 & $  58.6$ & $ -3.6$ &  8.2 & 2 \\
N Nor 2007     & V390 Nor   & 2007   6  15 & $ 338.6$ & $  2.2$ &  9.1 & 2 \\
N Sgr 2007     & V5558 Sgr  & 2007   4  14 & $  13.0$ & $  1.0$ &  8.0 & 2 \\
N Oph 2007     & V2615 Oph  & 2007   3  19 & $   5.3$ & $  4.0$ &  9.8 & 2 \\
N Cyg 2007     & V2467 Cyg  & 2007   3  15 & $  80.1$ & $  1.8$ &  6.7 & 2 \\
N Sco 2007\#2   & V1281 Sco  & 2007   2  19 & $ 349.4$ & $  5.2$ &  9.2 & 2 \\
N Sco 2007\#1   & V1280 Sco  & 2007   2   4 & $ 351.9$ & $  7.0$ &  3.9 & 2 \\
N Cen 2007     & V1065 Cen  & 2007   1  23 & $ 293.9$ & $  3.7$ &  8.5 & 2 \\
N Oph 2006\#2   & V2576 Oph  & 2006   4   6 & $ 356.5$ & $  5.6$ &  9.3 & 2 \\
RN Oph 1898    & RS Oph     & 2006   0   0 & $  21.1$ & $ 11.1$ &  4.8 & 2 \\
N Cyg 2006     & V2362 Cyg  & 2006   4   2 & $  87.4$ & $ -2.4$ &  8.2 & 2 \\
N Sgr 2006     & V5117 Sgr  & 2006   2  17 & $ 356.0$ & $ -5.6$ &  9.2 & 2 \\
N Cen 2005     & V1047 Cen  & 2005   8   1 & $ 306.4$ & $  1.3$ &  8.8 & 2 \\
N Sco 2005     & V1188 Sco  & 2005   7  25 & $ 355.8$ & $ -2.3$ &  8.0 & 2 \\
N Sgr 2005\#2   & V5116 Sgr  & 2005   7   4 & $   2.9$ & $ -6.4$ &  8.2 & 2 \\
N Sgr 2005\#1   & V5115 Sgr  & 2005   3  28 & $   7.7$ & $ -3.7$ &  7.7 & 2 \\
N Nor 2005     & V382 Nor   & 2005   3  13 & $ 333.1$ & $ -0.2$ &  9.8 & 2 \\
N Pup 2004     & V574 Pup   & 2004  11  20 & $ 242.4$ & $ -1.9$ &  6.9 & 2 \\
N Sco 2004\#2   & V1187 Sco  & 2004   8   3 & $ 357.0$ & $  2.3$ &  9.9 & 2 \\
N Sgr 2004     & V5114 Sgr  & 2004   3  15 & $   5.0$ & $ -5.7$ &  8.4 & 2 \\
N Sgr 2003\#2   & V5113 Sgr  & 2003   9  17 & $   5.0$ & $ -3.4$ &  8.9 & 2 \\
N Sct 2003     & V475 Sct   & 2003   8  28 & $  25.2$ & $ -3.4$ &  8.9 & 2 \\
N Sgr 2003\#1   & V4745 Sgr  & 2003   4  25 & $   2.3$ & $-12.0$ &  7.4 & 2 \\
N Cru 2003     & DZ Cru     & 2003   0   0 & $ 299.4$ & $  3.0$ &  9.2 & 2 \\
N Sgr 2002\#3   & V4743 Sgr  & 2002   9  20 & $  14.1$ & $-11.9$ &  5.0 & 2 \\
N Sgr 2002\#2   & V4742 Sgr  & 2002   9  15 & $   5.6$ & $ -1.1$ &  8.0 & 2 \\
N Oph 2002     & V2540 Oph  & 2002   1  24 & $   9.9$ & $  9.3$ &  9.2 & 2 \\
RN Nor 1920    & IM Nor     & 2002   1   3 & $ 327.5$ & $  3.0$ &  9.0 & 2 \\
N Sgr 2001\#3   & V4740 Sgr  & 2001   9   5 & $   2.0$ & $ -4.5$ &  7.0 & 2 \\
N Sgr 2001\#2   & V4739 Sgr  & 2001   8  26 & $   3.1$ & $ -7.3$ &  7.6 & 2 \\
N Cyg 2001\#2   & V2275 Cyg  & 2001   8  18 & $  88.9$ & $  1.5$ &  6.6 & 2 \\
N Sgr 2001\#1   & V4643 Sgr  & 2001   2  24 & $   3.7$ & $ -0.1$ &  8.1 & 2 \\
RN Aql 1917    & CI Aql     & 2000   4  30 & $  32.5$ & $ -0.4$ &  8.8 & 2 \\
RN Sco 1863    & U Sco      & 1999   2  25 & $ 359.1$ & $ 23.0$ &  7.6 & 2 \\
N Aql 1999\#2   & V1494 Aql  & 1999  12   1 & $  41.0$ & $ -4.7$ &  5.0 & 2 \\
N Cir 1999     & DD Cir     & 1999   8  23 & $ 311.1$ & $ -7.5$ &  7.5 & 2 \\
N Aql 1999\#1   & V1493 Aql  & 1999   7  13 & $  45.9$ & $  2.2$ & 10.0 & 2 \\
N Vel 1999     & V382 Vel   & 1999   5  22 & $ 283.8$ & $  6.5$ &  2.7 & 2 \\
N Sgr 1999     & V4444 Sgr  & 1999   4  25 & $   4.4$ & $ -3.1$ &  7.7 & 2 \\
N Mus 1998     & LZ Mus     & 1998  12  29 & $ 297.0$ & $ -2.2$ &  9.5 & 2 \\
N Sco 1998     & V1142 Sco  & 1998  10  21 & $ 359.3$ & $ -2.9$ &  6.9 & 2 \\
RN Oph 1900    & V2487 Oph  & 1998   6  15 & $   7.0$ & $  8.0$ &  9.5 & 2 \\
N Sgr 1998     & V4633 Sgr  & 1998   3  22 & $   6.1$ & $ -5.7$ &  7.4 & 2 \\
N Sco 1997     & V1141 Sco  & 1997   6   5 & $   0.1$ & $ -2.1$ &  8.5 & 2 \\
N Cru 1996     & CP Cru     & 1996   8  26 & $ 297.9$ & $  2.2$ &  9.2 & 2 \\
N Cas 1995     & V723 Cas   & 1995   8  24 & $ 125.0$ & $ -8.8$ &  7.1 & 2 \\
N Cen 1995     & V888 Cen   & 1995   2  23 & $ 304.3$ & $  3.0$ &  7.6 & 2 \\
N Aql 1995     & V1425 Aql  & 1995   2   7 & $  34.3$ & $ -3.3$ &  7.5 & 2 \\
N Cir 1995     & BY Cir     & 1995   1  27 & $ 315.8$ & $ -2.1$ &  7.2 & 2 \\
N Oph 1994     & V2313 Oph  & 1994   6   1 & $   7.5$ & $  7.3$ &  7.5 & 2 \\
N Sgr 1994\#2   & V4362 Sgr  & 1994   5  16 & $  15.6$ & $ -3.1$ &  8.0 & 2 \\
N Sgr 1994\#1   & V4332 Sgr  & 1994   2  24 & $  14.3$ & $ -8.9$ &  8.0 & 2 \\
N Cas 1993     & V705 Cas   & 1993  12   7 & $ 113.6$ & $ -4.1$ &  5.8 & 2 \\
N Lup 1993     & HY Lup     & 1993   9  19 & $ 318.5$ & $  9.0$ &  8.0 & 2 \\
N Sgr 1993     & V4327 Sgr  & 1993   8  14 & $   3.2$ & $ -4.8$ &  8.0 & 2 \\
N Aql 1993     & V1419 Aql  & 1993   5  13 & $  36.8$ & $ -4.1$ &  7.7 & 2 \\
N Oph 1993     & V2295 Oph  & 1993   4  14 & $   2.7$ & $  7.2$ &  9.0 & 2 \\
N Sgr 1992\#3   & V4171 Sgr  & 1992  10  13 & $  11.1$ & $ -3.5$ &  7.5 & 2 \\
N Sgr 1992\#2   & V4169 Sgr  & 1992   7   9 & $   5.2$ & $ -6.5$ &  7.7 & 2 \\
N Sco 1992     & V992  Sco  & 1992   5  22 & $ 344.2$ & $ -1.3$ &  7.3 & 2 \\
N Cyg 1992     & V1974 Cyg  & 1992   2  19 & $  89.1$ & $  7.9$ &  4.2 & 2 \\
N Sgr 1992\#1   & V4157 Sgr  & 1992   2  13 & $   6.8$ & $ -2.1$ &  7.0 & 2 \\
N Pup 1991     & V351  Pup  & 1991  12  27 & $ 252.4$ & $ -0.7$ &  6.4 & 2 \\
N Sgr 1991     & V4160 Sgr  & 1991   7  29 & $   0.5$ & $ -6.7$ &  7.0 & 2 \\
N Oph 1991\#2   & V2290 Oph  & 1991   4  11 & $   7.4$ & $  5.2$ &  9.3 & 2 \\
N Oph 1991\#1   & V2264 Oph  & 1991   4  11 & $   0.0$ & $  6.8$ &  9.8 & 2 \\
N Her 1991     & V838 Her   & 1991   3  24 & $  43.3$ & $  6.7$ &  5.0 & 2 \\
RN Sco 1937    & V745 Sco   & 1989   7  29 & $ 357.8$ & $ -3.8$ &  9.7 & 2 \\
RN Sgr 1962    & V3890 Sgr  & 1990   4  29 & $   9.2$ & $ -6.4$ &  8.6 & 2 \\
N Sco 1989     & V977 Sco   & 1989   8  17 & $ 358.4$ & $ -2.3$ & 10.0 & 2 \\
N Oph 1988     & V2214 Oph  & 1988   4  10 & $ 356.4$ & $  6.5$ &  8.5 & 2 \\
RN CrA 1949    & V394 CrA   & 1987   7  28 & $ 352.8$ & $ -7.7$ &  7.0 & 2 \\
N Vul 1987     & QV Vul     & 1987  11  15 & $  53.9$ & $  7.0$ &  7.0 & 2 \\
N Her 1987     & V827 Her   & 1987   1  25 & $  45.8$ & $  8.6$ &  7.4 & 2 \\
N And 1986     & OS And     & 1986  12   5 & $ 106.0$ & $-12.1$ &  6.3 & 3 \\
N Cen 1986     & V842 Cen   & 1986  11  22 & $ 317.1$ & $  3.6$ &  4.6 & 3 \\
N Cyg 1986     & V1819 Cyg  & 1986   8   4 & $  71.4$ & $  4.0$ &  8.7 & 3 \\
RN Oph 1898    & RS Oph     & 1985   1  28 & $  21.1$ & $ 11.1$ &  5.4 & 3 \\
N Aql 1985     & V1680 Aql  & 1985   0   0 & $  45.8$ & $  3.6$ &  9.7 & 3 \\
N Vul 1984\#2   & QU Vul     & 1984  12  18 & $  68.5$ & $ -6.0$ &  5.6 & 3 \\
N Aql 1984     & V1378 Aql  & 1984  12   2 & $  39.1$ & $ -3.9$ & 10.0 & 3 \\
N Vul 1984\#1   & PW Vul     & 1984   7  27 & $  61.1$ & $  5.2$ &  6.4 & 3 \\
N Nor 1983     & V341 Nor   & 1983   9  19 & $ 330.9$ & $ -1.1$ &  9.4 & 3 \\
N Ser 1983     & MU Ser     & 1983   2  22 & $  14.1$ & $  5.6$ &  7.7 & 3 \\
N Sgr 1983     & V4121 Sgr  & 1983   2  13 & $   4.0$ & $ -3.4$ &  9.5 & 3 \\
N Mus 1983     & GQ Mus     & 1983   1  18 & $ 297.1$ & $ -4.6$ &  7.2 & 3 \\
N Sgr 1982     & V4077 Sgr  & 1982  10   4 & $   8.2$ & $ -7.9$ &  8.0 & 3 \\
N Aql 1982     & V1370 Aql  & 1982   1  27 & $  38.8$ & $ -5.9$ &  6.0 & 3 \\
N CrA 1981     & V693 CrA   & 1981   4   2 & $ 358.8$ & $-14.0$ &  7.0 & 3 \\
N Sgr 1980     & V4065 Sgr  & 1980  10  28 & $   8.6$ & $ -3.7$ &  9.0 & 3 \\
RN Sco 1863    & U Sco      & 1979   6  24 & $ 359.1$ & $ 23.0$ &  8.7 & 3 \\
N Vul 1979     & PU Vul     & 1979   4   5 & $  62.6$ & $ -8.5$ &  9.0 & 3 \\
N Cyg 1978     & V1668 Cyg  & 1978   9  10 & $  90.8$ & $ -6.8$ &  6.7 & 3 \\
N Ser 1978     & LW Ser     & 1978   3   5 & $  14.2$ & $  6.8$ &  8.3 & 3 \\
N Sgr 1977     & V4021 Sgr  & 1977   3  27 & $  11.2$ & $ -7.4$ &  7.5 & 3 \\
N Sge 1977     & HS Sge     & 1977   1   7 & $  54.5$ & $ -1.9$ &  7.0 & 3 \\
N Vul 1976     & NQ Vul     & 1976  10  21 & $  55.4$ & $  1.3$ &  6.0 & 3 \\
N Oph 1976     & V2104 Oph  & 1976   9  23 & $  38.2$ & $ 15.9$ &  5.3 & 3 \\
N Cyg 1975     & V1500 Cyg  & 1975   8  29 & $  89.8$ & $ -0.1$ &  1.9 & 3 \\
N Sgr 1975     & V3889 Sgr  & 1975   7  13 & $   2.5$ & $ -1.7$ &  8.4 & 3 \\
N Sct 1975     & V373 Sct   & 1975   6  15 & $  27.8$ & $ -3.7$ &  7.1 & 3 \\
N Sgr 1975\#2   & V3964 Sgr  & 1975   6   8 & $  11.1$ & $  5.5$ &  6.0 & 3 \\
N Per 1974     & V400 Per   & 1974  11   9 & $ 145.7$ & $ -9.6$ &  7.8 & 3 \\
N Sgr 1976     & V3888 Sgr  & 1974  10   6 & $  10.4$ & $  5.4$ &  6.5 & 3 \\
N Cen 1973     & V812 Cen   & 1973   4  23 & $ 306.1$ & $  6.4$ &  9.0 & 3 \\
N Cep 1971     & IV Cep     & 1971   7  10 & $  99.6$ & $ -1.6$ &  7.0 & 3 \\
N Sct 1970     & V368 Sct   & 1970   7  31 & $  25.6$ & $ -2.1$ &  6.9 & 3 \\
N Sgr 1970     & V3645 Sgr  & 1970   7  29 & $  15.8$ & $ -4.5$ &  8.0 & 3 \\
N Cyg 1970     & V1330 Cyg  & 1970   6   8 & $  78.4$ & $ -5.5$ &  7.5 & 3 \\
N Aql 1970     & V1229 Aql  & 1970   4  14 & $  40.5$ & $ -5.4$ &  6.7 & 3 \\
N Ser 1970     & FH Ser     & 1970   2  13 & $  32.9$ & $  5.8$ &  4.5 & 3 \\
N Sgr 1969     & V2572 Sgr  & 1969   7   8 & $   2.6$ & $ -9.9$ &  6.5 & 3 \\
N Vul 1968\#2   & LU Vul     & 1968  10  14 & $  64.3$ & $  2.0$ &  9.5 & 3 \\
N Vul 1968\#1   & LV Vul     & 1968   4  15 & $  63.3$ & $  0.8$ &  5.2 & 3 \\
RN Oph 1898    & RS Oph     & 1967  10  12 & $  21.1$ & $ 11.1$ &  4.9 & 3 \\
RN Pyx 1890    & T Pyx      & 1967   1  15 & $ 256.6$ & $ 10.2$ &  6.4 & 3 \\
N Del 1967     & HR Del     & 1967   7   8 & $  63.4$ & $-14.0$ &  3.5 & 3 \\
N Oph 1967     & V2024 Oph  & 1967   7   7 & $   4.6$ & $  3.8$ &  9.5 & 3 \\
N CrA 1967     & V655 CrA   & 1967   6  30 & $ 358.7$ & $-10.2$ &  8.0 & 3 \\
N Aur 1964     & QZ Aur     & 1964  11   4 & $ 174.4$ & $ -0.7$ &  5.0 & 3 \\
N Sco 1964     & V825 Sco   & 1964   5  19 & $ 357.5$ & $ -2.6$ &  8.0 & 3 \\
N Pup 1963\#2   & HS Pup     & 1963  14  13 & $ 246.6$ & $ -1.4$ &  8.0 & 3 \\
N Her 1963     & V533 Her   & 1963   2   6 & $  69.2$ & $ 24.3$ &  3.0 & 3 \\
N Pup 1963\#1   & HZ Pup     & 1963   1  18 & $ 245.4$ & $  1.9$ &  7.7 & 3 \\
N Her 1960     & V446 Her   & 1960   3   7 & $  45.4$ & $  4.7$ &  3.0 & 3 \\
RN Oph 1898    & RS Oph     & 1958   7  14 & $  21.1$ & $ 11.1$ &  5.4 & 3 \\
N Oph 1957     & V972 Oph   & 1957   9  18 & $ 359.7$ & $  2.6$ &  8.0 & 3 \\
N UMi 1956     & RW UMi     & 1956   0   0 & $ 109.6$ & $ 33.2$ &  6.0 & 3 \\
N Sgr 1954\#1   & V1275 Sgr  & 1954   7   4 & $ 355.6$ & $ -5.9$ &  7.0 & 3 \\
N Oph 1954     & V908 Oph   & 1954   7   2 & $   0.0$ & $  4.9$ &  7.5 & 3 \\
N Cha 1953     & RR Cha     & 1953   4   8 & $ 304.3$ & $-18.9$ &  7.1 & 3 \\
N Oph 1952     & V906 Oph   & 1952   8  22 & $   5.1$ & $  8.5$ &  8.4 & 3 \\
N Sco 1952\#2   & V723 Sco   & 1952   8  10 & $ 355.6$ & $ -3.7$ &  9.8 & 3 \\
N Sco 1952\#1   & V722 Sco   & 1952   3  10 & $ 356.7$ & $ -2.6$ &  9.4 & 3 \\
N Sgr 1952\#1   & V1175 Sgr  & 1952   2  21 & $   1.4$ & $ -6.3$ &  7.0 & 3 \\
N Sgr 1951     & V1172 Sgr  & 1951   3   7 & $   8.8$ & $  4.0$ &  9.0 & 3 \\
N Sco 1950\#3   & V721 Sco   & 1950   9   3 & $ 355.9$ & $ -1.7$ &  8.0 & 3 \\
N Sco 1950\#2   & V720 Sco   & 1950   8   7 & $ 355.8$ & $ -4.1$ &  7.5 & 3 \\
N Sco 1950\#1   & V719 Sco   & 1950   7  20 & $ 355.7$ & $ -2.6$ &  9.8 & 3 \\
N Lac 1950     & DK Lac     & 1950   1  23 & $ 105.2$ & $ -5.4$ &  5.0 & 3 \\
N Sct 1949     & EU Sct     & 1949   8   5 & $  30.1$ & $ -2.8$ &  8.4 & 3 \\
RN CrA 1949    & V394 CrA   & 1949   3  23 & $ 352.8$ & $ -7.7$ &  7.5 & 3 \\
N Cyg 1948     & V465 Cyg   & 1948   6   2 & $  71.9$ & $  4.8$ &  7.3 & 3 \\
N Ser 1948     & CT Ser     & 1948   4   9 & $  24.5$ & $ 47.6$ &  5.0 & 3 \\
N Sgr 1947     & V928 Sgr   & 1947   5  16 & $   4.5$ & $ -5.9$ &  8.5 & 3 \\
RN CrB 1866    & T CrB      & 1946   2   9 & $  42.4$ & $ 48.2$ &  3.2 & 3 \\
RN Sco 1863    & U Sco      & 1945   5  30 & $ 359.1$ & $ 23.0$ &  9.9 & 3 \\
N Aql 1945     & V528 Aql   & 1945   8  26 & $  36.7$ & $ -5.9$ &  7.0 & 3 \\
N Sgr 1945     & V1149 Sgr  & 1945   2  16 & $   4.6$ & $ -5.7$ &  7.4 & 3 \\
RN Pyx 1890    & T Pyx      & 1944  11  13 & $ 256.6$ & $ 10.2$ &  6.3 & 3 \\
N Sco 1944     & V696 Sco   & 1944   5   5 & $ 356.3$ & $ -4.1$ &  7.5 & 3 \\
N Sgr 1944     & V927 Sgr   & 1944   4  22 & $ 359.2$ & $ -6.0$ &  7.3 & 3 \\
N Aql 1943     & V500 Aql   & 1943   9   5 & $  47.6$ & $ -9.5$ &  6.1 & 3 \\
N Sgr 1943     & V1148 Sgr  & 1943   8   3 & $   6.9$ & $ -2.1$ &  8.0 & 3 \\
N Mon 1942     & KT Mon     & 1942  13   2 & $ 205.1$ & $ -3.3$ &  9.8 & 3 \\
N Pup 1942     & CP Pup     & 1942  11   9 & $ 252.3$ & $ -0.4$ &  0.5 & 3 \\
N Cyg 1942     & V450 Cyg   & 1942   9   8 & $  79.1$ & $ -6.5$ &  7.0 & 3 \\
N Sgr 1941     & V909 Sgr   & 1941   7  18 & $ 358.8$ & $-10.4$ &  6.8 & 3 \\
N Sco 1941     & V697 Sco   & 1941   3   9 & $ 354.1$ & $ -5.0$ &  8.0 & 3 \\
N Vel 1940     & CQ Vel     & 1940   4  19 & $ 271.8$ & $ -4.5$ &  8.9 & 3 \\
N Mon 1939     & BT Mon     & 1939  12  17 & $ 213.8$ & $ -2.6$ &  5.0 & 3 \\
N Sgr 1937     & V787 Sgr   & 1937   5  20 & $   1.1$ & $ -3.0$ &  9.8 & 3 \\
RN Sco 1863    & U Sco      & 1936   6  21 & $ 359.1$ & $ 23.0$ &  8.4 & 3 \\
N Aql 1936\#2   & V368 Aql   & 1936  10   7 & $  43.7$ & $ -4.3$ &  6.1 & 3 \\
N Sgr 1936     & V630 Sgr   & 1936  10   3 & $ 358.4$ & $ -6.6$ &  4.5 & 3 \\
N Aql 1936\#1   & V356 Aql   & 1936   9  18 & $  37.4$ & $ -4.9$ &  7.7 & 3 \\
N Lac 1936     & CP Lac     & 1936   6  18 & $ 102.1$ & $ -0.8$ &  2.1 & 3 \\
N Sgr 1936     & V732 Sgr   & 1936   6  10 & $   3.2$ & $ -0.8$ &  6.5 & 3 \\
N Her 1934     & DQ Her     & 1934  12  12 & $  73.2$ & $ 26.4$ &  1.3 & 3 \\
RN Oph 1898    & RS Oph     & 1933   8  12 & $  19.5$ & $ 11.0$ &  5.4 & 3 \\
N Sgr 1932     & V1905 Sgr  & 1932   6  14 & $   8.9$ & $ -7.3$ &  8.0 & 3 \\
N Cen 1931     & MT Cen     & 1931   5   9 & $ 294.4$ & $  2.3$ &  8.4 & 3 \\
N Sgr 1930     & V441 Sgr   & 1930   9  12 & $   7.9$ & $ -4.9$ &  8.0 & 3 \\
N Sgr 1928     & V1583 Sgr  & 1928   7  24 & $   8.8$ & $ -2.7$ &  8.9 & 3 \\
N Sco 1928     & KP Sco     & 1928   6  21 & $ 355.3$ & $ -2.5$ &  9.4 & 3 \\
N Tau 1927     & XX Tau     & 1927  11  18 & $ 187.1$ & $-11.7$ &  5.9 & 3 \\
N Sgr 1927     & V363 Sgr   & 1927   9  30 & $   9.2$ & $-16.4$ &  8.7 & 3 \\
N Aql 1927     & EL Aql     & 1927   7  30 & $  31.1$ & $ -2.2$ &  6.4 & 3 \\
N Sgr 1926     & FM Sgr     & 1926   7  16 & $   9.2$ & $ -2.9$ &  8.0 & 3 \\
N Sgr 1926     & KY Sgr     & 1926   6  14 & $   4.7$ & $ -1.3$ &  8.0 & 3 \\
N Aql 1925     & DO Aql     & 1925   9  14 & $  32.5$ & $-11.4$ &  8.7 & 3 \\
N Pic 1925     & RR Pic     & 1925   5  25 & $ 271.0$ & $-25.4$ &  1.0 & 3 \\
N Sgr 1924     & FL Sgr     & 1924   5  13 & $ 357.8$ & $ -5.0$ &  8.0 & 3 \\
N Sgr 1924     & GR Sgr     & 1924   4  30 & $   8.0$ & $ -5.1$ &  7.5 & 3 \\
N Sco 1922     & V707 Sco   & 1922   7  11 & $ 354.8$ & $ -3.8$ &  9.6 & 3 \\
RN Pyx 1890    & T Pyx      & 1920   4  11 & $ 256.6$ & $ 10.2$ &  6.9 & 3 \\
RN Nor 1920    & IM Nor     & 1920   7  27 & $ 327.5$ & $  3.0$ &  9.5 & 3 \\
N Cyg 1920     & V476 Cyg   & 1920   8  20 & $  87.4$ & $ 12.4$ &  2.0 & 3 \\
N Lyr 1919     & HR Lyr     & 1919  12   6 & $  59.6$ & $ 12.5$ &  6.5 & 3 \\
N Oph 1919     & V849 Oph   & 1919   8  20 & $  39.2$ & $ 13.5$ &  7.2 & 3 \\
N Sgr 1919     & V1017 Sgr  & 1919   3  11 & $   5.2$ & $ -8.8$ &  7.2 & 3 \\
N Aql 1918     & V603 Aql   & 1918   6   8 & $  33.2$ & $  0.8$ & -1.1 & 3 \\
N Mon 1918     & GI Mon     & 1918   2   4 & $ 221.7$ & $  5.4$ &  5.2 & 3 \\
RN Aql 1917    & CI Aql     & 1917   5  20 & $  32.5$ & $ -0.4$ &  8.7 & 3 \\
RN Sco 1863    & U Sco      & 1917   3   6 & $ 359.1$ & $ 23.0$ &  8.8 & 3 \\
N Sgr 1917     & BS Sgr     & 1917   7  17 & $   6.2$ & $ -6.9$ &  9.2 & 3 \\
N Oph 1917     & V840 Oph   & 1917   4  25 & $ 354.1$ & $  9.5$ &  5.5 & 3 \\
N Sgr 1914     & V1012 Sgr  & 1914   8  12 & $   1.1$ & $ -4.5$ &  8.0 & 3 \\
N Gem 1912     & DN Gem     & 1912   3  12 & $ 184.0$ & $ 14.7$ &  3.5 & 3 \\
N Lac 1910     & DI Lac     & 1910  12  30 & $ 103.1$ & $ -4.9$ &  4.6 & 3 \\
N Ara 1910     & OY Ara     & 1910   4   4 & $ 334.5$ & $ -3.4$ &  5.1 & 3 \\
N Sgr 1910     & V999 Sgr   & 1910   3  21 & $   3.8$ & $ -1.5$ &  8.0 & 3 \\
RN Sco 1863    & U Sco      & 1906   5  12 & $ 359.1$ & $ 23.0$ &  8.8 & 3 \\
N Sco 1906     & V711 Sco   & 1906   4  24 & $ 356.9$ & $ -4.0$ &  9.7 & 3 \\
N Aql 1905     & V604 Aql   & 1905   8  31 & $  31.0$ & $ -4.0$ &  7.6 & 3 \\
N Sgr 1905     & V1015 Sgr  & 1905   8  18 & $   0.3$ & $ -5.7$ &  6.5 & 3 \\
N Ser 1903     & X Ser      & 1903   5   0 & $  11.8$ & $ 32.4$ &  8.9 & 3 \\
N Gem 1903     & DM Gem     & 1903   3  16 & $ 185.1$ & $ 11.7$ &  4.8 & 3 \\
RN Pyx 1890    & T Pyx      & 1902   5   2 & $ 256.6$ & $ 10.2$ &  6.9 & 3 \\
N Pup 1902     & DY Pup     & 1902  11  19 & $ 244.9$ & $  5.0$ &  7.0 & 3 \\
RN Pyx 1890    & T Pyx      & 1902   0   0 & $ 256.6$ & $ 10.2$ &  6.5 & 3 \\
N Sco 1901     & V382 Sco   & 1901   9   2 & $ 355.8$ & $ -4.1$ &  9.5 & 3 \\
N Per 1901     & GK Per     & 1901   2  21 & $ 151.0$ & $-10.1$ &  0.2 & 3 \\
N Sgr 1900     & HS Sgr     & 1900   0   0 & $  12.1$ & $ -4.3$ & 10.0 & 3 \\
N Sgr 1900     & AT Sgr     & 1900   0   0 & $   5.0$ & $ -1.7$ & 10.0 & 3 \\

\enddata
\tablenotetext{a}{Peak magnitude when available, otherwise discovery magnitude}
\tablenotetext{b}{(1) {\tt http://asd.gsfc.nasa.gov/Koji.Mukai/novae/novae.html}; (2) {\tt http://www.cbat.eps.harvard.edu/nova\_list.html}; (3) {\tt http://projectpluto.com/galnovae/galnovae.htm}; (4) Duerbeck (1987)\hfill}

\end{deluxetable}